\documentclass[a4paper,fleqn,usenatbib]{mnras}
\usepackage{graphicx}  % needed for figures
\usepackage{dcolumn}   % needed for some tables
\usepackage{amssymb}   % for math
\usepackage{amsmath}   % for math
\usepackage{mathptmx}
\usepackage{xspace}
\usepackage{url}
\usepackage{bm}
\usepackage{footnote}
\usepackage{float}
\usepackage{color}
\def\reff@jnl#1{{\rm#1\/}}
\def\aj{\reff@jnl{AJ}}         % Astronomical Journal
\def\araa{\reff@jnl{ARA\&A}}      % Annual Review of Astron and Astrophys
\def\apj{\reff@jnl{ApJ}}        % Astrophysical Journal
\def\apjl{\reff@jnl{ApJ}}        % Astrophysical Journal, Letters
\def\apjs{\reff@jnl{ApJS}}       % Astrophysical Journal, Supplement
\def\aap{\reff@jnl{A\&A}}        % Astronomy and Astrophysics
\def\aapr{\reff@jnl{A\&A~Rev.}}     % Astronomy and Astrophysics Reviews
\def\aaps{\reff@jnl{A\&AS}}       % Astronomy and Astrophysics, Supplement
\def\mnras{\reff@jnl{MNRAS}}      % Monthly Notices of the RAS
\def\physrep{\reff@jnl{Physics Reports}}% Physics Reports
\def\prd{\reff@jnl{Phys.Rev.D}}     % Physical Review D
\def\prl{\reff@jnl{Phys.Rev.Lett}}   % Physical Review Letters
\def\pasp{\reff@jnl{PASP}}       % Publications of the ASP
\def\pasj{\reff@jnl{PASJ}}       % Publications of the ASJ
\def\nat{\reff@jnl{Nature}}       % Nature
\def\jcap{\reff@jnl{JCAP}}   %Journal of Cosmology and Astroparticle Physics
\def\memsai{\reff@jnl{MemSAI}} %Memorie della Societa Astronomica Italiana Supplement 
\def\na{\reff@jnl{New Astronomy}}       % New Astronomy

\newcommand{\simgt}{\lower.5ex\hbox{$\; \buildrel > \over \sim \;$}}
\newcommand{\simlt}{\lower.5ex\hbox{$\; \buildrel < \over \sim \;$}}

\def\Sref#1{$\S$\ref{#1}\xspace}

\def\Fref#1{Figure~\ref{#1}\xspace}

\def\Tref#1{Table~\ref{#1}\xspace}

\def\Eref#1{Eqn.~\ref{#1}\xspace}

\def\Aref#1{Appendix~\ref{#1}\xspace}
\def\Cref#1{Chapter~\ref{#1}\xspace}

\def\gvec{\boldsymbol{\gamma}}

\def\tvec{\boldsymbol{\theta}}
\def\lvec{\boldsymbol{\ell}}

\def\btwo{1.12\pm0.19}
\def\bthree{0.97\pm0.15}
\def\bfour{1.38\pm0.39}
\def\bfive{1.45\pm0.56}

\newcommand{\chihway}[1]{{\color{black}#1}}

\def\anl{Argonne National Laboratory, 9700 South Cass Avenue, Lemont, IL 60439, USA}
\def\upenn{Department of Physics and Astronomy, University of Pennsylvania, Philadelphia, PA 19104, USA}
\def\ethz{Department of Physics, ETH Zurich, Wolfgang-Pauli-Strasse 16, CH-8093 Zurich, Switzerland}
\def\ports{Institute of Cosmology \& Gravitation, University of Portsmouth, Portsmouth, PO1 3FX, UK}
\def\ucl{Department of Physics \& Astronomy, University College London, Gower Street, London, WC1E 6BT, UK}
\def\bnl{Brookhaven National Laboratory, Bldg 510, Upton, NY 11973, USA}
\def\fermilab{Fermi National Accelerator Laboratory, P. O. Box 500, Batavia, IL 60510, USA}
\def\stanford{Department of Physics, Stanford University, 382 Via Pueblo Mall, Stanford, CA 94305, USA}
\def\kipac{Kavli Institute for Particle Astrophysics \& Cosmology, P. O. Box 2450, Stanford University, Stanford, CA 94305, USA}
\def\slac{SLAC National Accelerator Laboratory, Menlo Park, CA 94025, USA}
\def\ifae{Institut de F\'{\i}sica d'Altes Energies, Universitat Aut\`onoma de Barcelona, E-08193 Bellaterra, Barcelona, Spain}
\def\ieec{Institut de Ci\`encies de l'Espai, IEEC-CSIC, Campus UAB, Facultat de Ci\`encies, Torre C5 par-2, 08193 Bellaterra, Barcelona, Spain}
\def\ccap{Center for Cosmology and Astro-Particle Physics, The Ohio State University, Columbus, OH 43210, USA}
\def\ohio{Department of Physics, The Ohio State University, Columbus, OH 43210, USA}
\def\manchester{Jodrell Bank Center for Astrophysics, School of Physics and Astronomy, University of Manchester, Oxford Road, Manchester, M13 9PL, UK}
\def\cambridgekavli{Kavli Institute for Cosmology, University of Cambridge, Madingley Road, Cambridge CB3 0HA, UK}
\def\cambridge{Institute of Astronomy, University of Cambridge, Madingley Road, Cambridge CB3 0HA, UK}
\def\paris{CNRS, UMR 7095, Institut d'Astrophysique de Paris, F-75014, Paris, France}
\def\lina{Laborat\'orio Interinstitucional de e-Astronomia - LIneA, Rua Gal. Jos\'e Cristino 77, Rio de Janeiro, RJ - 20921-400, Brazil}
\def\on{Observat\'orio Nacional, Rua Gal. Jos\'e Cristino 77, Rio de Janeiro, RJ - 20921-400, Brazil}
\def\texas{George P. and Cynthia Woods Mitchell Institute for Fundamental Physics and Astronomy, and Department of Physics and Astronomy,
Texas A\&M University, College Station, TX 77843,  USA}

\def\michigan{Department of Physics, University of Michigan, Ann Arbor, MI 48109, USA}
\def\michiganastro{Department of Astronomy, University of Michigan, Ann Arbor, MI 48109, USA}

\def\maxplanck{Max Planck Institute for Extraterrestrial Physics, Giessenbachstrasse, 85748 Garching, Germany}

\def\lmu{Faculty of Physics, Ludwig-Maximilians-Universitaet, Scheinerstr. 1, 81679 Muenchen, Germany}
\def\ctio{Cerro Tololo Inter-American Observatory, National Optical Astronomy Observatory, Casilla 603, La Serena, Chile}
\def\aao{Australian Astronomical Observatory, North Ryde, NSW 2113, Australia}

\def\jpl{Jet Propulsion Laboratory, California Institute of Technology, 4800 Oak Grove Dr., Pasadena, CA 91109, USA}
\def\ciemat{Centro de Investigaciones Energ\'eticas, Medioambientales y Tecnol\'ogicas (CIEMAT), Madrid, Spain}
\def\uiuc{Department of Physics, University of Illinois, 1110 W. Green St., Urbana, IL 61801, USA}
\def\ncsa{National Center for Supercomputing Applications, 1205 West Clark St., Urbana, IL 61801, USA}

\def\jpl{Jet Propulsion Laboratory, California Institute of Technology, 4800 Oak Grove Dr., Pasadena, CA 91109, USA}
\def\sussex{Department of Physics and Astronomy, Pevensey Building, University of Sussex, Brighton, BN1 9QH, UK}
\def\cluster{Excellence Cluster Universe, Boltzmannstr.\ 2, 85748 Garching, Germany}
\def\barcelona{Instituci\'o Catalana de Recerca i Estudis Avan\c{c}ats, E-08010 Barcelona, Spain}

\def\anl{Argonne National Laboratory, 9700 South Cass Avenue, Lemont, IL 60439, USA}
\def\ucl{Department of Physics \& Astronomy, University College London, Gower Street, London, WC1E 6BT, UK}
\def\rhodes{Department of Physics and Electronics, Rhodes University, PO Box 94, Grahamstown, 6140, South Africa}
\def\kicp{Kavli Institute for Cosmological Physics, University of Chicago, Chicago, IL 60637, USA}
\def\princeton{Department of Astrophysical Sciences, Princeton University, Peyton Hall, Princeton, NJ 08544, USA}
\def\lbnl{Lawrence Berkeley National Laboratory, 1 Cyclotron Road, Berkeley, CA 94720, USA}
\def\berkeley{Department of Astronomy, University of California, Berkeley,  501 Campbell Hall, Berkeley, CA 94720, USA}
\def\southhampton{School of Physics and Astronomy, University of Southampton,  Southampton, SO17 1BJ, UK}
\def\sorbonne{Sorbonne Universit\'es, UPMC Univ Paris 06, UMR 7095, Institut d'Astrophysique de Paris, F-75014, Paris, France}
\def\fisica{Departamento de F\'{\i}sica Matem\'atica,  Instituto de F\'{\i}sica, Universidade de S\~ao Paulo,  CP 66318, CEP 05314-970, S\~ao Paulo, SP,  Brazil}
\def\damtp{Centre for Theoretical Cosmology, DAMTP, University of Cambridge, Wilberforce Road, Cambridge CB3 0WA, UK}

\begin{document}

\title[Galaxy bias from DES SV]{Galaxy bias from the Dark Energy Survey Science Verification data: 
\\ combining galaxy density maps and weak lensing maps}

\author{DES}

\author[C.~Chang, A.~Pujol, E.~Gazta\~{n}aga, A.~Amara et al.]
{C.~Chang,$^{1*}$ 
A.~Pujol,$^{2}$ 
E.~Gazta\~{n}aga,$^{2}$ 
A.~Amara,$^{1}$ 
A.~R\'efr\'egier,$^{1}$ 
\newauthor
D.~Bacon,$^{3}$
M.~R.~Becker,$^{4,5}$
C.~Bonnett,$^{6}$
J.~Carretero,$^{2,6}$
F.~J.~Castander,$^{2}$
M.~Crocce,$^{2}$ 
\newauthor
P.~Fosalba,$^{2}$
T.~Giannantonio,$^{7,8,9}$
W.~Hartley,$^{1}$
M.~Jarvis,$^{10}$
T.~Kacprzak,$^{1}$
\newauthor
A.~J.~Ross,$^{11}$
E.~Sheldon,$^{12}$
M.~A.~Troxel,$^{13}$
V.~Vikram,$^{14}$
J.~Zuntz,$^{13}$ 
T. M. C.~Abbott,$^{15}$ 
\newauthor
F.~B.~Abdalla,$^{16,17}$
S.~Allam,$^{18}$
J.~Annis,$^{18}$
A.~Benoit-L{\'e}vy,$^{16,19,20}$
E.~Bertin,$^{19,20}$
\newauthor
D.~Brooks,$^{16}$
E.~Buckley-Geer,$^{18}$
D.~L.~Burke,$^{5,21}$
D.~Capozzi,$^{3}$
A.~Carnero~Rosell,$^{22,23}$
\newauthor
M.~Carrasco~Kind,$^{24,25}$
C.~E.~Cunha,$^{5}$
C.~B.~D'Andrea,$^{3,26}$
L.~N.~da Costa,$^{22,23}$
\newauthor
S.~Desai,$^{27,28}$
H.~T.~Diehl,$^{18}$
J.~P.~Dietrich,$^{27,28}$
P.~Doel,$^{16}$
T.~F.~Eifler,$^{10,29}$
J.~Estrada,$^{18}$
\newauthor
A.~E.~Evrard,$^{30,31}$
B.~Flaugher,$^{18}$
J.~Frieman,$^{18,32}$
D.~A.~Goldstein,$^{33,34}$
D.~Gruen,$^{5,21,28,35}$
\newauthor
R.~A.~Gruendl,$^{24,25}$
G.~Gutierrez,$^{18}$
K.~Honscheid,$^{11,36}$
B.~Jain,$^{10}$
D.~J.~James,$^{15}$
\newauthor
K.~Kuehn,$^{37}$
N.~Kuropatkin,$^{18}$
O.~Lahav,$^{16}$
T.~S.~Li,$^{38}$
M.~Lima,$^{22,39}$
J.~L.~Marshall,$^{38}$
\newauthor
P.~Martini,$^{11,36}$
P.~Melchior,$^{40}$
C.~J.~Miller,$^{30,31}$
R.~Miquel,$^{6,41}$
J.~J.~Mohr,$^{27,28,35}$
\newauthor
R.~C.~Nichol,$^{3}$
B.~Nord,$^{18}$
R.~Ogando,$^{22,23}$
A.~A.~Plazas,$^{29}$
K.~Reil,$^{21}$
A.~K.~Romer,$^{42}$
\newauthor
A.~Roodman,$^{5,21}$
E.~S.~Rykoff,$^{5,21}$
E.~Sanchez,$^{43}$
V.~Scarpine,$^{18}$
M.~Schubnell,$^{31}$
\newauthor
I.~Sevilla-Noarbe,$^{24,43}$
R.~C.~Smith,$^{15}$
M.~Soares-Santos,$^{18}$
F.~Sobreira,$^{18,22}$
E.~Suchyta,$^{10}$
\newauthor
M.~E.~C.~Swanson,$^{25}$
G.~Tarle,$^{31}$
D.~Thomas,$^{3}$
A.~R.~Walker$^{15}$ \\
\vspace{0.02in} \\
(all affiliations at the end of paper)
\vspace{0.02in} \\
$^{*}$e-mail address: chihway.chang@phys.ethz.ch\\
}
 
\date\today
\maketitle

\begin{abstract}
We measure the redshift evolution of galaxy bias for a magnitude-limited galaxy sample by combining the 
galaxy density maps and weak lensing shear maps for a $\sim$116 deg$^{2}$ area of the Dark Energy 
Survey (DES) Science Verification data. This method was first developed in \citet{Amara2012} 
and later re-examined in a companion paper \citep{Pujol2016} with rigorous simulation tests and 
analytical treatment of tomographic measurements. In this work we apply this method to the  
DES SV data and measure the galaxy bias for a $i<22.5$ galaxy sample. 
We find the galaxy bias and 1$\sigma$ error bars in 4 photometric redshift bins to be 
\chihway{
$\btwo$ ($z = 0.2 - 0.4$), $\bthree$ ($z = 0.4 - 0.6$), $\bfour$ ($z = 0.6 - 0.8$), and 
$\bfive$ ($z = 0.8 - 1.0$).
These measurements are consistent at the 2$\sigma$ level with measurements on the same dataset using 
galaxy clustering and cross-correlation of galaxies with CMB lensing, with most of the redshift bins consistent 
within the 1$\sigma$ error bars.} In addition, our method provides the 
only $\sigma_{8}$-independent constraint among the three. We forward-model the main observational 
effects using mock galaxy catalogs by including shape noise, photo-$z$ errors and masking effects. We 
show that our bias measurement from the data is consistent with that expected from simulations. With the 
forthcoming full DES data set, we expect this method to provide additional constraints on the galaxy bias 
measurement from more traditional methods. Furthermore, in the process of our measurement, we build up 
a 3D mass map that allows further exploration of the dark matter distribution and its relation to galaxy 
evolution.
\end{abstract}


\section{Introduction}

Galaxy bias is one of the key ingredients for describing our observable Universe. In a concordance 
$\Lambda$CDM model, galaxies form at overdensities of the dark matter distribution, suggesting the 
possibility of simple relations between the distribution of galaxies and dark matter. This particular relation is 
described by a galaxy bias model \citep{Kaiser1984}. Galaxy bias bridges the observable Universe 
of galaxies with the underlying dark matter. For a full review of literature on galaxy bias, we refer the readers 
to \citet{Eriksen2015} and references therein. 

Observationally, several measurement techniques exist for constraining galaxy bias. The most common approach 
is to measure galaxy bias through the 2-point correlation function (2PCF) of galaxies \citep{Blake2008, Simon2009, Cresswell2009, 
Coupon2012, Zehavi2011}. 
Counts-in-cells (CiC) is another method where the higher moments of the galaxy probability density function (PDF) 
are used to constrain galaxy bias \citep{Blanton2000, Wild2005, Swanson2008}. 
Alternatively, one can combine galaxy clustering with measurements from gravitational 
lensing, which probes the total (baryonic and dark) matter distribution. Such measurements include combining 
galaxy clustering with galaxy-galaxy lensing \citep{Simon2007, Jullo2012, Mandelbaum2013} and lensing of the 
cosmic microwave background (CMB) \citep{Schneider1998, Giannantonio2016}. The method we present in this work also belongs to this class. 

With ongoing and upcoming large galaxy surveys 
(the Hyper SuprimeCam\footnote{\url{www.naoj.org/Projects/HSC}}, 
the Dark Energy Survey\footnote{\url{www.darkenergysurvey.org}}, 
the Kilo Degree Survey\footnote{\url{kids.strw.leidenuniv.nl}}, 
the Large Synoptic Survey Telescope\footnote{\url{www.lsst.org}}, 
the Euclid mission\footnote{\url{sci.esa.int/euclid}}, 
the Wide-Field Infrared Survey Telescope\footnote{\url{wfirst.gsfc.nasa.gov}}), 
statistical uncertainties on the galaxy bias measurements will decrease significantly. It is thus interesting to explore 
alternative and independent options of measuring galaxy bias. Such measurements would be powerful tests 
for systematic uncertainties and break possible degeneracies.  

In this paper, we present a new measurement of the redshift-dependent galaxy bias from the Dark Energy Survey 
(DES) Science Verification (SV) data using a novel method. Our method relies on the cross-correlation between 
weak lensing shear and galaxy density maps to constrain galaxy bias. The method naturally combines the 
power of galaxy surveys and weak lensing measurements in a way that only weakly depends on assumptions 
of the cosmological parameters. In addition, the method involves building up a high-resolution 3D mass map in 
the survey volume which is interesting for studies of the dark matter distribution at the map level. The relation 
between the galaxy sample and the mass map also provides information for studies of galaxy evolution.

The analysis in this paper closely follows \citet[][hereafter A12]{Amara2012} 
and \citet[][hereafter Paper I]{Pujol2016}.  A12 applied this method to COSMOS and zCOSMOS data and discussed 
different approaches for constructing the galaxy density map and galaxy bias. Paper I carried out a series of simulation 
tests to explore the regime of the measurement parameters where the method is consistent with 2PCF measurements, 
while introducing alternative approaches to the methodology. 
Building on these two papers, this work applies the method to the DES SV data, demonstrating the first constraints 
with this method using photometric data. Simulations are used side-by-side with data to ensure that each step 
in the data analysis is robust. In particular, we start with the same set of ``ideal'' simulations used in Paper I and 
gradually degrade until they match the data by including noise, photometric redshift errors, and masking effects. 

The paper is organized as follows. In \Sref{sec:background} we overview the basic principles of our measurement 
method. 
In \Sref{sec:data} we introduce the data and simulations used in this work. 
The analysis and results are presented in \Sref{sec:results}, first with a series of simulation tests and then with the 
DES SV data. We also present a series of systematics tests here. 
In \Sref{sec:comparison} we compare our measurements with bias measurements on the same data 
set using different approaches. We conclude in \Sref{sec:conclusion}.  

\section{Background theory}
\label{sec:background}

\subsection{Linear galaxy bias}
\label{sec:bias}
In this work we follow Paper I, where the overdensities of galaxies $\delta_{g}$ is linearly related to 
the overdensities of dark matter $\delta$ at some given smoothing scale $R$, or 
\begin{equation}
\delta_{g}(z, R) = b(z, R) \delta(z, R). 
\label{eq:bias}
\end{equation}  
We define $\delta \equiv \frac{\rho-\bar{\rho}}{\bar{\rho}}$, where $\rho$ is the dark matter density and 
$\bar{\rho}$ is the mean dark matter density at a given redshift. $\delta_{g}$ is defined similarly, with $\rho$ 
replaced by $\rho_{g}$, the number density of galaxies. $b$ can depend on galaxy properties such as luminosity, 
color and type \citep{Swanson2008, Cresswell2009}. This definition is often referred to as the ``local bias'' 
model. According to \citet{Manera2011}, at sufficiently large scales ($\gtrsim 40$ Mpc/h comoving distance), $b(z,R)$ in \Eref{eq:bias} 
is consistent with galaxy bias defined through the 2PCF of dark matter ($\xi_{dm}$) and galaxies ($\xi_{g}$). That is, 
the following equation holds,
\begin{equation}
\xi_{g}(r) = \langle \delta_{g}(\boldsymbol{r_{0}})\delta_{g}(\boldsymbol{r_{0}}+\boldsymbol{r}) \rangle = b^{2}\langle \delta (\boldsymbol{r_{0}})\delta(\boldsymbol{r_{0}}+\boldsymbol{r}) \rangle = b^{2} \xi_{dm}(r),
\label{eq:xi}
\end{equation}
where $\boldsymbol{r_{0}}$ and $\boldsymbol{r_{0}}+\boldsymbol{r}$ are two positions on the sky separated by vector $\boldsymbol{r}$. The angle bracket $\langle \rangle$ averages over all pairs of positions on the sky separated by distance $|\boldsymbol{r}|\equiv r$. 
Our work will be based on scales in this regime.

\subsection{Weak Lensing}
\label{sec:wl}
Weak lensing refers to the coherent distortion, or ``shear'' of galaxy images caused by large-scale cosmic 
structures between these galaxies and the observer. 
Weak lensing probes directly the total mass instead of a proxy 
of the total mass (e.g. stellar mass, gas mass). For a detailed review of the theoretical background of weak 
lensing, see e.g. \citet{Bartelmann2001}.

The main weak lensing observable is the complex shear $\gvec = \gamma_{1}+i\gamma_{2}$, which is estimated 
by the measured shape of galaxies. The cosmological shear signal is much weaker than the intrinsic galaxy shapes. 
The uncertainty in the shear estimate due to this intrinsic galaxy shape is referred to as ``shape noise'', and is often 
the largest source of uncertainty in lensing measurements.
Shear can be converted to convergence, $\kappa$, a scalar field that directly 
measures the projected mass. 
The convergence at a given position $\tvec$ on the sky can be expressed as
\begin{equation}   
\kappa(\tvec, p_{s}) = \int_{0}^{\infty} d\chi \: q(\chi, p_{s}) \delta(\tvec, \chi),
\label{eq:kappa2}
\end{equation}
where $q(\chi,p_{s})$ is the lensing weight
\begin{equation}
q(\chi, p_{s}) \equiv \frac{3 H_{\rm 0}^{2} \Omega_{\rm m} \chi}{2c^{2}a(\chi)}\int_{\chi}^{\infty} d\chi_{s} \frac{\chi_{s}-\chi}{\chi_{s}}p_{s}(\chi_{s}).
\end{equation}
Here, $\chi$ is the comoving distance, $\Omega_{\rm m}$ 
is the total matter density of the Universe today normalised by the critical density today, 
$H_{\rm 0}$ is the Hubble constant today, and $a$ is the scale factor. $p_{s}(\chi)$ is the 
normalized redshift distribution of the ``source'' galaxy sample where the lensing quantities ($\gvec$ or $\kappa$) 
are measured. In the simple case of a single source redshift plane at $\chi_{s}$, $p_{s}$ is 
a delta function and the lensing weight becomes 
\begin{equation}
q(\chi, \chi_{s}) \equiv \frac{3 H_{\rm 0}^{2} \Omega_{\rm m}}{2c^{2}a(\chi)}\frac{\chi(\chi_{s}-\chi)}{\chi_{s}}.
\end{equation}

In the flat-sky approximation, conversion between $\gvec$ and $\kappa$ in Fourier space follows 
\citep[][KS conversion]{Kaiser1993}:
\begin{equation}
\tilde{\kappa}(\lvec)-\tilde{\kappa_{0}} = D^{*}(\lvec)\tilde{\gvec}(\lvec); \;\; \tilde{\gvec}(\lvec)-\tilde{\gvec_{0}} = D(\lvec)\tilde{\kappa}(\lvec),
\label{eq:kappa2gamma}
\end{equation}
where ``$\tilde{X}$'' indicates the Fourier transform of the field $X$, $\lvec$ is the spatial frequency, 
$\tilde{\kappa_{0}}$ and $\tilde{\gvec_{0}}$ are small constant offsets which cannot be reconstructed and are 
often referred to as the ``mass-sheet degeneracy''. $D$ is a combination of second moments of $\lvec$:
\begin{equation}
D(\lvec)=\frac{\ell_{1}^{2}-\ell_{2}^{2} +i2\ell_{1}\ell_{2}}{|\lvec|^{2}}.
\end{equation}

In this work we follow the implementation of \Eref{eq:kappa2gamma} as described 
in \citet{Vikram2015} and \citet{Chang2015c} to construct $\kappa$ and $\gvec$ maps as needed.

\subsection{$\kappa_{g}$: a convergence template from galaxies}
\label{sec:kappag}
Following the same approach as A12 and Paper I, we now define $\kappa_{g}$ by substituting $\delta$ 
with $\delta_{g}$ in \Eref{eq:kappa2}, or 
\begin{equation}   
\kappa_{g}(\tvec, p_{s}) = \int_{0}^{\infty} d\chi \: q(\chi, p_{s}) \delta_{g}(\tvec, \chi).
\label{eq:kappa3}
\end{equation} 
Physically, $\kappa_{g}$ is a ``template'' for the convergence $\kappa$. In particular, in the case of a constant 
galaxy bias $b$, where $\delta_{g}=b \delta$ everywhere, \Eref{eq:kappa3} trivially gives 
$\kappa_{g}=b\kappa$. The relation between $\kappa$, $\kappa_{g}$ 
and $b$ in the case of redshift-dependent galaxy bias (\Eref{eq:bias}) becomes more complicated. This 
requires the introduction of the ``partial'' $\kappa_{g}$, or $\kappa'_{g}$ below. Alternatively, one can 
adopt the approach used in A12 and include a parametrized galaxy bias model in constructing 
$\kappa_{g}$. 

To construct $\kappa'_{g}$, instead of integrating over all foreground ``lens'' galaxies in \Eref{eq:kappa3}, we only consider the 
part of the template contributed by a given lens sample. This gives
\begin{align}   
\kappa'_{g}(\tvec, \phi', p_{s}) &= \int_{0}^{\infty} d\chi \: q(\chi, p_{s}) \phi'(\chi) \delta_{g}(\tvec, \chi) \notag \\
&= \int_{0}^{\infty} d\chi \: q(\chi, p_{s}) \phi'(\chi) \left(\frac{\rho_{g}(\tvec, \chi)}{\bar{\rho_{g}}(\chi)}-1\right),
\label{eq:kappa4}
\end{align} 
where $\phi'(\chi)$ is the radial selection function of the lens sample of interest. $\rho_{g}$ is the 
number of galaxies per unit volume and $\bar{\rho_{g}}$ is the mean of $\rho_{g}$ at a given redshift. 
$\phi'(\chi)$ is different from $p'(\chi)$ in Eqn. 20 of Paper I only by a normalization: 
$\int d\chi p'(\chi) = 1$, while $\phi'(\chi)$ integrates to a length, which is the origin of the $\Delta \chi'$ in Eqn. 
20 in Paper I. We choose to use $\phi'(\chi)$ here to facilitate the derivation later, but note that \Eref{eq:kappa8} 
below is fully consistent with Eqn. 20 in Paper I.    
Similarly we define also a partial $\kappa$ field, which we will later use in 
\Sref{sec:bias_cl_wl}, 
\begin{equation}   
\kappa'(\tvec, \phi', p_{s}) = \int_{0}^{\infty} d\chi \: q(\chi, p_{s}) \phi'(\chi) \delta(\tvec, \chi).
\label{eq:kappa6}
\end{equation} 

In practice, when constructing $\kappa'_{g}$, we assume a fixed source redshift $\bar{\chi}_{s}$ and 
take the mean lensing weight $\bar{q}'$ and $\bar{\rho}_{g}$ outside the integration of \Eref{eq:kappa4}.
This approximation holds in the case where $q$ and $\bar{\rho_{g}}$ are slowly varying over the extent 
of $\phi'$, which is true for the intermediate redshift ranges we focus on. We have
\begin{equation}
\kappa'_{g}(\tvec, \phi', \bar{\chi}_{s})  \approx \Delta \chi' \bar{q}'(\bar{\chi}_{s}) \left( \frac{\int_{0}^{\infty} d\chi \phi'(\chi)\rho_{g}(\tvec, \chi)}{\bar{\rho_{g}}\Delta \chi'} - 1\right),
\label{eq:kappa7}
\end{equation}
where 
\begin{equation}
\Delta \chi' = \int_{0}^{\infty} d\chi \phi'(\chi).
\end{equation}
We further simplify the expression by defining the partial 2D surface density $\Sigma'$ and $\bar{\Sigma}'$, where
\begin{equation}
\Sigma' = \int_{0}^{\infty}d\chi \phi'(\chi) \rho_{g}(\tvec, \chi), \; \bar{\Sigma}' = \int_{0}^{\infty}d\chi \phi'(\chi) \bar{\rho}_{g}. 
\end{equation} 
\Eref{eq:kappa7} then becomes
\begin{equation}
\kappa'_{g}(\tvec, \phi', \bar{\chi}_{s})  \approx \Delta \chi' \bar{q}'(\bar{\chi}_{s}) \left( \frac{\Sigma'(\tvec)}{\bar{\Sigma}'(\tvec)} - 1\right),
\label{eq:kappa8}
\end{equation}
which is what we measure as described in \Sref{sec:procedure}. 
 
\subsection{Bias estimation from the galaxy density field and the weak lensing field}
\label{sec:bias_cl_wl}

The information of galaxy bias can be extracted through the cross- and auto-correlation of 
the $\kappa$ and $\kappa'_{g}$ fields. (In the case of constant bias, we can replace $\kappa'_{g}$ 
by $\kappa_{g}$ in all equations below.) Specifically, we calculate 
\begin{equation}
b' = \frac{\langle \kappa'_{g}\kappa'_{g} \rangle}{\langle \kappa'_{g}\kappa\rangle} = \frac{\langle \kappa'_{g}(\tvec,\phi', \bar{\chi}_{s}) \kappa'_{g}(\tvec,\phi', \bar{\chi}_{s}) \rangle}{\langle \kappa'_{g}(\tvec,\phi', \bar{\chi}_{s})\kappa(\tvec, p_{s})\rangle},
\label{eq:zerolag}
\end{equation}
where $\langle \rangle$ represents a zero-lag correlation between the two fields in the brackets, 
averaged over a given aperture $R$. We can write for the most general case, 
\begin{equation}
\langle \kappa_{A} \kappa_{B} \rangle = \frac{4 \pi}{\pi^{2}R^{4}} \int_{0}^{R} dr_{1} r_{1}  \int_{0}^{R} dr_{2} r_{2}
\int_{0}^{\pi} d\eta \omega_{AB} (\Theta),
\end{equation}
where $\kappa_{A}$ and $\kappa_{B}$ can be any of the following: $(\kappa,\kappa', \kappa_{g}, \kappa'_{g})$, 
$\Theta^{2}=r_{1}^{2}+r_{2}^{2}-2r_{1}r_{2}\cos \eta$, and $\omega_{AB} (\Theta)$ is the projected two-point angular 
correlation function between the two fields, defined 
\begin{equation}
\omega_{AB}(\Theta) = \int_{0}^{\infty} d\chi_{A}\int_{0}^{\infty} d\chi_{B} q_{A} q_{B} \phi'_{A} \phi'_{B} \xi_{\kappa_{A}\kappa_{B}}(r),
\end{equation}
where $q_{A}$ ($q_{B}$) and $\phi'_{A}$ ($\phi'_{B}$) are the lensing weight and lens redshift selection function associated 
with the fields $\kappa_{A}$ ($\kappa_{B}$). $\xi_{\kappa_{A}\kappa_{B}}(r)$ is the 3D two-point correlation function. 
In the case of $\kappa_{A}=\kappa_{B}=\kappa$, $\xi_{\kappa_{A}\kappa_{B}}$ reduces to $\xi_{dm}$ in 
\Eref{eq:xi}.

\begin{figure}
  \begin{center}
   \includegraphics[scale=0.48]{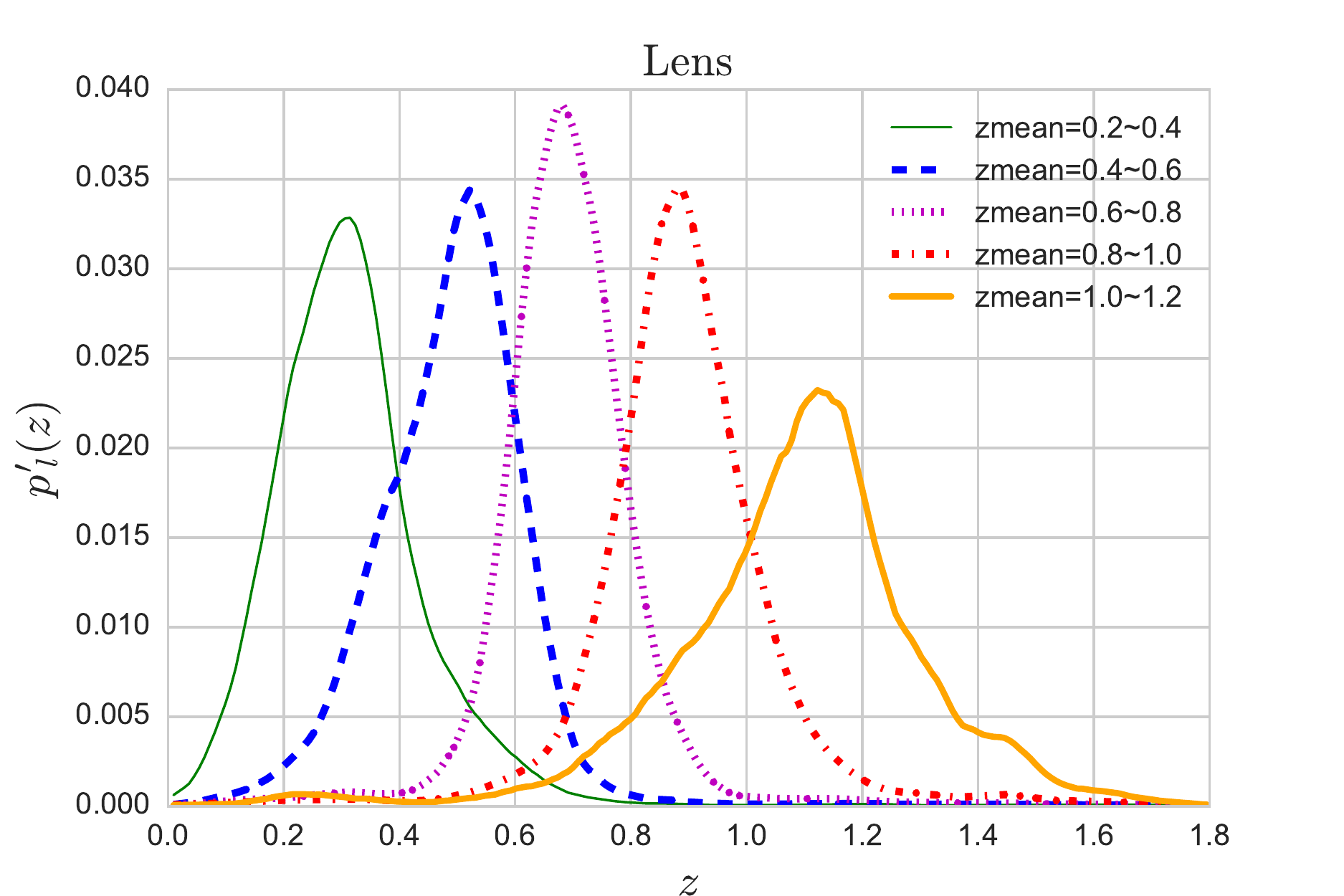}  
   \includegraphics[scale=0.48]{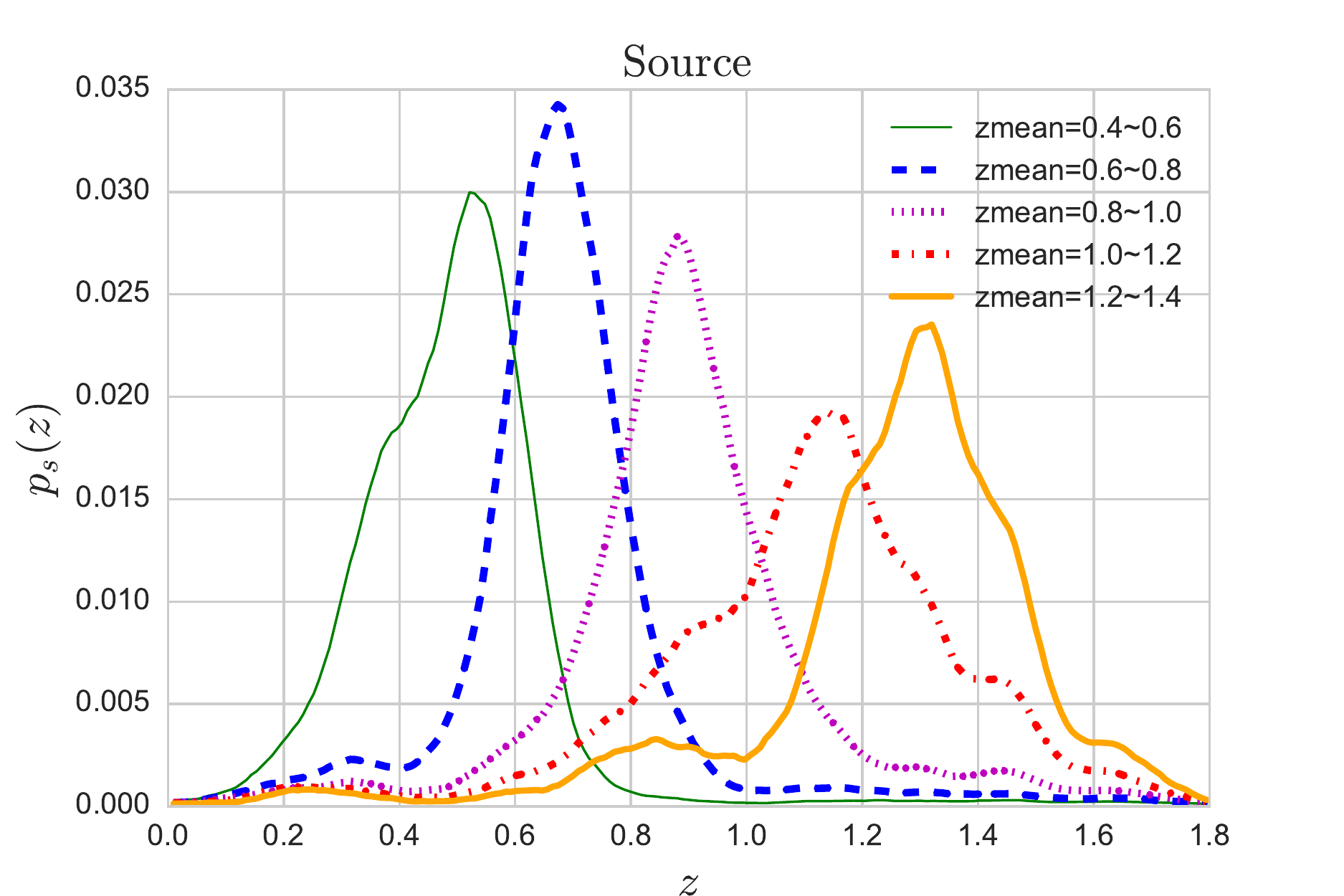}  
  \caption{Normalized redshift distribution of the lens (top) and source (bottom) samples as estimated from the 
  photo-$z$ code \textsc{Skynet}. Each curve represents the stacked PDF for all galaxies in the photo-z bin 
  determined by $ z_{\mathrm{mean}}$ as listed in the labels.   }
  \label{fig:z_dist}
  \end{center}
\end{figure}

For infinitely thin redshift bins, or constant bias, $b'$ in \Eref{eq:zerolag} directly measures the galaxy bias $b$ 
of the lens. However, once the lens and source samples span a finite redshift range (see eg. \Fref{fig:z_dist}), 
$b'$ is a function of the source and lens distribution and is different from $b$ by some factor $f(\phi', p_{s})$, so that 
\begin{equation}
b'=f(\phi', p_{s})b.
\label{eq:bfb}
\end{equation} 
Note that $f$ can be determined if $b(z)$ is known. Since we have $b(z)=1$ for the case of dark matter, we can 
calculate $f$ by calculating $b'$ and setting $b(z)=1$,  or
\begin{equation}
f(\phi', p_{s}) = \frac{\langle \kappa' \kappa' \rangle}{\langle \kappa' \kappa\rangle} =  \frac{\langle \kappa'(\tvec,\phi', \bar{\chi}_{s}) \kappa'(\tvec,\phi', \bar{\chi}_{s}) \rangle}{\langle \kappa'(\tvec,\phi', \bar{\chi}_{s})\kappa(\tvec, p_{s})\rangle},
\label{eq:zerolag2}
\end{equation}
where $\kappa'$ is defined in \Eref{eq:kappa6} and follows the same assumptions in \Eref{eq:kappa8}, 
where the lensing weight depends on only the mean distance to the source sample $\bar{\chi}_{s}$. 
$f$ here corresponds to $f_{2}$ in Eqn. 26 in Paper I. \chihway{\Tref{tab:f} shows an example of the $f$ 
values calculated from the data. }

\begin{table}
\begin{center}
\caption{\chihway{$f$ factor (\Eref{eq:zerolag2}) calculated from data. $f$ depends on the specific 
sample that is used. In this table we list numbers only for the main measurement in 
\Sref{sec:data_measurement}, where 
the \textsc{ngmix} shear catalog and the \textsc{Skynet} photo-z catalog is used.}}
\begin{tabular}{lcccc}
Source                             & \multicolumn{4}{c}{Lens redshift} \\ 
redshift  &  $0.2-0.4$ & $0.4-0.6$ & $0.6-0.8$ & $0.8-1.0$  \\  \hline
$0.4-0.6$ & 0.61 & -- & -- &--    \\   
$0.6-0.8$  & 0.61 & 0.58 &-- & --   \\  
$0.8-1.0$  &  0.61& 0.59 & 0.67  & --  \\   
$1.0-1.2$  &  0.62 & 0.60  &0.72  & 0.53  \\  
\end{tabular}
\label{tab:f}
\end{center}
\end{table}

\chihway{We use a slightly different estimator for $b'$ compared to \Eref{eq:zerolag} in practice. Combined with \Eref{eq:bfb}, our 
estimator for galaxy bias is: 
\begin{equation}
b = \frac{1}{\mu},
\label{eq:mu}
\end{equation}
\begin{equation}
\mu = f\frac{\langle \gamma'_{\alpha, g}\gamma'_{\alpha}\rangle}{\langle \gamma'_{\alpha, g} \gamma'_{\alpha, g} \rangle - \langle \gamma^{'N}_{\alpha, g} \gamma^{'N}_{\alpha, g} \rangle},
\label{eq:zerolag3}
\end{equation}
with $\alpha=1,2$ referring to the two components of $\gvec$. }

\chihway{Here we replaced $\kappa'$ by $\gamma'_{\alpha}$, which is possible since the two quantities are 
interchangeable through \Eref{eq:kappa2gamma}. The main reason to work with $\gamma'_{\alpha}$ is that in our data 
set, $\gamma'_{\alpha}$ is much noisier compared to the $\kappa'_{g}$ due to the presence of the shape noise, 
therefore converting $\gamma'_{\alpha}$ to $\kappa'_{\alpha}$ would be suboptimal to converting 
$\kappa'_{g}$ to $\gamma'_{\alpha,g}$. This choice depends somewhat on the specific data quality at hand. 
In addition, the term $\langle \gamma^{'N}_{\alpha, g} \gamma^{'N}_{\alpha, g} \rangle$ is introduced to account for 
the shot noise arising from the finite number of galaxies in the galaxy density field (see also Paper I). The term is 
calculated by randomizing the galaxy positions when calculating $\gamma'_{\alpha,g}$. 
Finally, since $\langle \gamma'_{\alpha, g}\gamma'_{\alpha}\rangle$ is noisy and can become close to zero, 
measuring directly the inverse of \Eref{eq:zerolag3}
results in a less stable and biased estimator. Therefore, we measure the inverse-bias, $\mu$, throughout the 
analysis and only take the inverse at the very end to recover the galaxy bias $b$. This approach is 
similar to that used in A12. We show in \Aref{sec:estimator} the results using $b$ instead of $\mu$ as 
our main estimator.}

The measurement from this method would depend on assumptions of the cosmological model in the construction 
of $\kappa'_{g}$ and the calculation of $f$. Except for the literal linear dependence on $H_{\rm 0}$$\Omega_{\rm m}$, 
due to the ratio nature of the measurement, most other parameters tend to cancel out. Within the current constraints 
from \textit{Planck}, the uncertainty in the cosmological parameters affect the measurements at the percent level, 
which is well within the measurement errors ($>10\%$). All cosmological parameters used in the calculation of this 
work are consistent with the simulations described in \Sref{sec:sims}.

\subsection{Multiple source-lens samples}
\label{sec:multiple_bin}
Whereas \Eref{eq:mu} and \Eref{eq:zerolag3} describes how we measure galaxy bias for one source sample and 
one lens sample, in practice multiple different samples of lenses and the sources are involved. We define several 
source and lens samples, or ``bins'', based on their photometric redshift (photo-$z$), with the lens samples labeled 
by $i$ and the source samples labeled by $j$. \chihway{We use the notation $\mu^{\alpha}_{ij}$ to represent the 
inverse-bias measured with $\gamma_{\alpha}$ using the source bin $j$ and lens bin $i$. }

\chihway{Our estimate of the galaxy bias in each lens redshift bin $i$ is calculated by combining 
$\mu^{\alpha}_{ij}$ estimates from the two components of shear and all source redshift bins $j$. To 
combine these, we consider a least-square fit to the following model 
\begin{equation}
D_{i} = \bar{\mu}_{i}M,
\end{equation}
where $D= \{\mu^{\alpha}_{ij}\}$ is the data vector containing all the measurements $\mu^{\alpha}_{ij}$ 
of galaxy bias in this lens bin $i$ (including measurement from the two shear components and possibly 
multiple source bins), $\bar{\mu}_{i}$ is the combined inverse-bias in each bin $i$ we wish to fit for, and 
$M$ is a 1D array of the same length as $D_{i}$ with all elements being 1. Our final estimate of inverse-bias 
for redshift bin $i$, $\bar{\mu_{i}}$, and its uncertainty $\sigma(\bar{\mu_{i}})$ are: 
\begin{equation}
\bar{\mu}_{i} = M_{i}^{T}C_{i}^{-1}D_{i}[M_{i}^{T}C_{i}^{-1}M_{i}]^{-1},
\label{eq:weight_mean}
\end{equation} 
\begin{equation}
\sigma(\bar{\mu}_{i})^{2} = (M_{i}^{T}C_{i}^{-1}M_{i})^{-1},
\label{eq:weight_std}
\end{equation} 
where $C_{i}^{-1}$ is the unbiased inverse covariant matrix \citep{Hartlap2007} between all $\mu^{\alpha}_{ij}$ 
measurements, estimated by Jack-Knife (JK) resampling: 
\begin{equation}
C_{i}^{-1} = \tau \: \mathrm{Cov}^{-1}[D_{i}],  %covariance matrix
\end{equation}
where $\tau=(N-\nu-2)/(N-1)$. $N$ is the number of JK samples, and $\nu$ is the dimension of $C_{i}$.
Note that the matrix inversion of $C_{i}$ becomes unstable when the measurements $\mu^{\alpha}_{ij}$ are highly 
correlated. This is the case in the noiseless simulations. For the noisy simulations and data, however, it does 
not affect the results. 
The galaxy bias and its uncertainty is then
\begin{equation}
\bar{b}_{i} = \frac{1}{\bar{\mu}_{i}} 
\end{equation}
and
\begin{equation}
\sigma^{2}(\bar{b}_{i}) = \frac{\sigma^{2}(\bar{\mu}_{i})}{\bar{\mu}^{2}_{i}}.
\end{equation}}
  
\chihway{The uncertainty estimated through JK resampling does not account for cosmic variance and its 
coupling with the mask geometry. In \Sref{sec:sim_test}, we further include the uncertainty from cosmic 
variance using simulations.} 

\section{Data and simulations}
\label{sec:data}

In this section we describe the data and simulation used in this work. 
We use the DES SV data collected using the Dark Energy Camera \citep{Flaugher2015} from 
November 2012 to February 2013 and that have been processed through the Data Management 
pipeline described in \citet{Ngeow2006, Sevilla2011, 
Desai2012, Mohr2012}. Individual images are stacked, objects are detected and their 
photometric/morphological properties are measured using the software packages \textsc{SCAMP} 
\citep{Bertin2006}, \textsc{SWARP} \citep{Bertin2002}, \textsc{PSFEx} \citep{Bertin2011} and 
\textsc{SExtractor} \citep{Bertin1996}. The final product, the SVA1 Gold 
catalog\footnote{\url{http://des.ncsa.illinois.edu/releases/sva1}} 
is the foundation of all catalogs described below. We use a $\sim$116.2 deg$^{2}$ subset of the data in the 
South Pole Telescope East (SPT-E) footprint, which is the largest contiguous region in the SV dataset.   
This data set is also used in other DES weak lensing and large-scale structure analyzes \citep{Vikram2015, 
Chang2015c, Becker2015, DES2015, Crocce2016, Giannantonio2016}.

\subsection{Photo-$z$ catalog}
\label{sec:photoz}

The photo-$z$ of each galaxy is estimated through the \textsc{Skynet} code \citep{Graff2014}. 
\textsc{Skynet} is a machine learning algorithm that has been extensively tested in \citet{Sanchez2014} and 
\citet{Bonnett2015} to perform well in controlled simulation tests. To test the robustness of our results, we also 
carry out our main analysis using two other photo-$z$ codes which were tested in \citet{Sanchez2014} and 
\citet{Bonnett2015}: 
\textsc{BPZ} \citep{Benitez2000}, and \textsc{TPZ} \citep{Carrasco2013, 
Carrasco2014}. We discuss in \Sref{sec:systematics} the results from these different photo-$z$ codes. 

The photo-$z$ codes output a PDF for each galaxy describing 
the probability of the galaxy being at redshift $z$. We first use the mean of the PDF, $ z_{\mathrm{mean}}$ to separate 
the galaxies into redshift bins, and then use the full PDF to calculate \Eref{eq:zerolag2}. 
In \Fref{fig:z_dist}, we show the normalized redshift distribution 
for each lens and source bin as defined below. 

\subsection{Galaxy catalog}
To generate the $\kappa_{g}$ maps, we use the same ``Benchmark'' sample used in \citet{Giannantonio2016} 
and \citet{Crocce2016}. This is a magnitude-limited galaxy sample at $18<i<22.5$ derived from the 
SVA1 Gold catalog with additional cleaning with color, region, and star-galaxy classification cuts 
\citep[see][for full details of this sample]{Crocce2016}. The final 
area is $\sim 116.2$ square degrees with an average galaxy number density of 5.6 per arcmin$^{2}$. 
Six redshift bins were used from $ z_{\mathrm{mean}}=0.0$ to $ z_{\mathrm{mean}}=1.2$ with $\Delta  z_{\mathrm{mean}} = 0.2$. 
The magnitude-limited sample is constructed by using only the sky regions with limiting magnitude 
deeper than $i=22.5$, where the limiting magnitude is estimated by modelling the survey depth as a 
function of magnitude and magnitude errors \citep{Rykoff2015}. Various systematics tests on the Benchmark 
has been performed in \citet{Crocce2016} and \citet{Leistedt2015}.

\subsection{Shear catalog}
\label{sec:data_shear}
Two shear catalogs are available for the DES SV data based on two independent shear measurement 
codes \textsc{ngmix} \citep{Sheldon2014} and \textsc{im3shape} \citep{Zuntz2013}. Both catalogs 
have been tested rigorously in \citet{Jarvis2015} and have been shown to pass the requirements 
on the systematic uncertainties for the SV data. Our main analysis is based on \textsc{ngmix} due to 
its higher effective number density of galaxies (5.7 per arcmin$^{2}$ compared to 3.7 per arcmin$^{2}$ 
for \textsc{im3shape}). However we check in \Sref{sec:sys_im3shape} that 
both catalogs produce consistent results. We adopt the selection cuts recommended in \citet{Jarvis2015} 
for both catalogs. This galaxy sample is therefore consistent with the other DES SV measurements in 
e.g., \citet{Becker2015, DES2015}. Similar to these DES SV papers, we perform all our measurements 
on a blinded catalog (for details of the blinding procedure, see \citet{Jarvis2015}), and only un-blind 
when the analysis is finalized. 

$\gamma_{1}$  and $\gamma_{2}$ maps are generated from the shear catalogs for five redshift bins 
between $ z_{\mathrm{mean}}=0.4$ and $ z_{\mathrm{mean}}=1.4$ with $\Delta  z_{\mathrm{mean}}=0.2$. 
Note part of the highest redshift bin lies outside of the recommended photo-$z$ selection according to 
\citet{Bonnett2015} 
($ z_{\mathrm{mean}}=0.3-1.3$). We discard the highest bin in the final analysis due to low signal-to-noise 
(see \Sref{sec:data_measurement}), but for future work, however, it would be 
necessary to validate the entire photo-$z$ range used.

\subsection{Mask}
\label{sec:mask}
Two masks are used in this work. First, we apply a common mask to all maps used in this work, we 
will refer this mask as the \textit{``map mask''}. The mask is constructed by re-pixelating the 
$i>22.5$ depth map into the coarser (flat) pixel grid of 5$\times$5 arcmin$^{2}$ we use to 
construct all maps (see \Sref{sec:procedure}). 
The depth mask has a much higher resolution ($nside=4096$ \texttt{Healpix} map) than this grid, which 
means some pixels in the new grid will be partially masked in the original \texttt{Healpix} grid. We 
discard pixels in the new grid with more than half of the area masked in the \texttt{Healpix} grid. The 
remaining partially masked pixels causes effectively a $\sim3\%$ increase in the total area. The partially 
masked pixels will be taken into account later when generating $\kappa_{g}$ (we scale the mean number 
of galaxy per pixel by the appropriate pixel area). We also discard pixels without any source galaxies. 

\begin{figure}
  \begin{center}
   \includegraphics[scale=0.35]{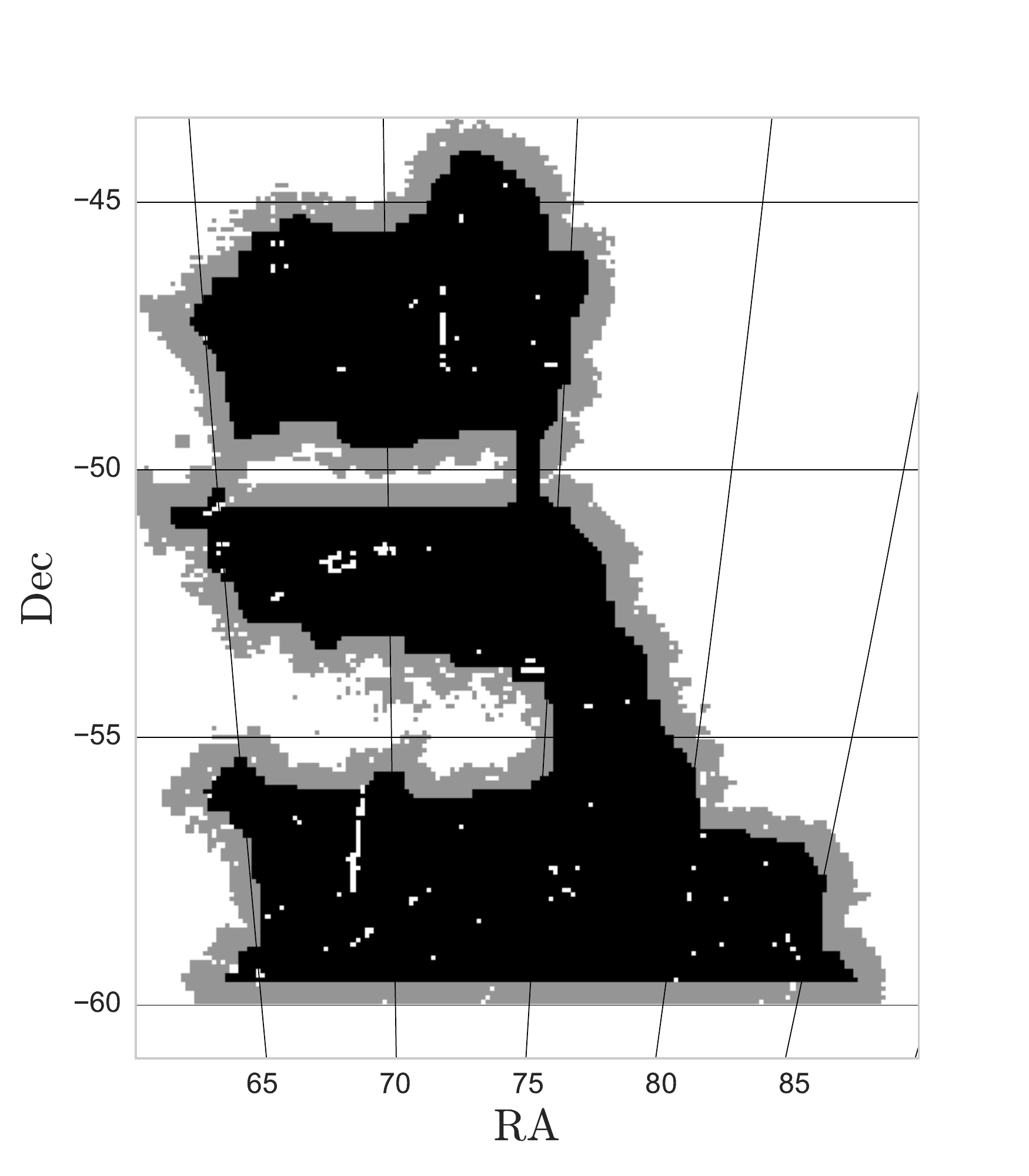}  
  \caption{Mask used in this work. The black region shows where the galaxy bias is calculated (the 
  \textit{bias mask}). The black+grey map region is where all maps are made (the \textit{map mask}).}
  \label{fig:mask}
  \end{center}
\end{figure}

Pixels on the edges of our mask will be affected by the smoothing we apply 
to the maps. In addition, when performing the KS conversion, the mask can affect our results. We 
thus define a second \textit{``bias mask''}, where we start from the \textit{map mask} and further mask 
pixels that are closer than half a smoothing scale away from any masked pixels except for holes 
smaller than 1.5 pixels\footnote{The reason for not apodizing the small masks is that it would reduce 
significantly the region unmasked and thus the statistical power of our measurement. We have tested 
in simulations that the presence of these small holes do not affect our final measurements.. We 
consider only pixels surviving the \textit{bias mask} when estimating 
galaxy bias. \Fref{fig:mask} shows both masks used in this work.}

\subsection{Simulations}
\label{sec:sims}
In this work we use the same mock galaxy catalog from the MICE 
simulations\footnote{\url{http://cosmohub.pic.es/}} \citep{Fosalba2015, Crocce2015, Fosalba2015b} 
which is described in detail in Paper I. MICE adopts the $\Lambda$CDM cosmological parameters: 
$\Omega_{\rm m}=0.25$, $\sigma_{8}=0.8$, $n_{s}=0.95$, $\Omega_{\rm b}=0.044$, $\Omega_{\rm \Lambda}=0.75$ 
and $h=0.7$. The galaxy catalogue has been generated according to a Halo Occupation Distribution
(HOD) and a SubHalo Abundance Matching (SHAM) prescription described in \citet{Carretero2015}.
The main tests were done with the region $0^{\circ}<$RA$<30^{\circ}$, $0^{\circ}<$Dec$<30^{\circ}$, 
while we use a larger region ($0^{\circ}<$RA$<90^{\circ}$, $0^{\circ}<$Dec$<30^{\circ}$) to estimate 
the effect from cosmic variance. We use the following properties for each 
galaxy in the catalog -- position on the sky (RA, Dec), redshift ($z$), apparent magnitude in the $i$ 
band, and weak lensing shear ($\gvec$). % and convergence ($\kappa$).   

In addition, we incorporate shape noise and masking effects that are matched to the data. For shape 
noise, we draw randomly from the ellipticity distribution in the data and add linearly to the true shear in the mock 
catalog to yield ellipticity measurements for all galaxies in the mock catalog. We also make sure that 
the source galaxy number density is matched between simulation and data in each redshift bin. For the mask, we 
simply apply the same mask from the data to the simulations. Note that the un-masked simulation area is  
$\sim8$ times larger than the data, thus applying the mask increases the statistical uncertainty. 

Finally, to investigate the effect of photo-$z$ uncertainties, we add a Gaussian photo-$z$ error to each 
MICE galaxy according to its true redshift. The standard deviation of the Gaussian uncertainty follows 
$\sigma(z) = 0.03(1+z)$. 
This model for the photo-$z$ error is simplistic, but since we use this set of photo-$z$ simulations mainly 
to test our algorithm (the calculation of $f$ in \Eref{eq:zerolag2}), we believe a simple model will serve its purpose.

We note that the larger patch of MICE 
simulation used in this work ($\sim 30\times 30$ square degrees) is of the order of what is expected 
for the first year of DES data ($\sim$2,000 degree square and $\sim1$ magnitude shallower). Thus, the 
simulation measurements shown in this work also serves as a rough forecast for our method applied on 
the first year of DES data.

\section{Analysis and results}
\label{sec:results}

\subsection{Procedure}
\label{sec:procedure}

Before we describe the analysis procedure, it is helpful to have a mental picture of a 3D cube in 
RA, Dec and $z$. The $z$-dimension is illustrated in \Fref{fig:z_dist}, with a coarse resolution of five 
redshift bins for both lenses and sources. Each lens and source sample is then collapsed into 2D maps 
in the RA/Dec-dimension. For each source bin, we can only constrain the galaxy bias using the lens bins 
at the foreground of this source bin. That is, for the highest source redshift bin there are five corresponding 
lens bins, and for the lowest source redshift bin there is only one lens bin. The analysis is carried out in the 
following steps. 

First, we generate all the necessary maps for the measurement: $\gamma_{1}$, $\gamma_{2}$ maps for 
each source redshift bin $j$, and $\gamma'_{1,g}$, $\gamma'_{2,g}$, $\gamma^{'N}_{1,g}$, and $\gamma^{'N}_{2,g}$ 
maps for each lens bin $i$ and source bin $j$. \chihway{We generate random maps ($\gamma^{'N}_{1,g}$, 
$\gamma^{'N}_{2,g}$) for the calculation of $\langle \gamma^{'N}_{\alpha, g} \gamma^{'N}_{\alpha, g} \rangle$ 
in \Eref{eq:zerolag3}.} All maps are generated using a sinusoidal projection at a 
reference RA of 71$^{\circ}$ and 5 arcmin square pixels on the projected plane. These maps are then smoothed 
by a 50 arcmin boxcar filter while the \textit{map mask} is applied. The chosen pixel and smoothing scales are 
based on tests described in Paper I. 
For a given source bin, the value of each pixel in the $\gamma_{1}$ and $\gamma_{2}$ maps is simply the 
weighted mean of the shear measurements in the area of that pixel. The weights reflect 
the uncertainties in the shear measurements in the data, while we set all weights to 1 in the simulations. 
For a given lens bin, the pixel values of the $\gamma'_{1,g}$, $\gamma'_{2,g}$ maps are calculated through 
\Eref{eq:kappa8}, where $\Sigma'$ is the number of galaxies in that pixel, and $\bar{\Sigma}'$ is the 
mean number of galaxies per pixel in that lens bin.
For each combination of lens-source bins, we calculate \chihway{$\mu^{\alpha}_{ij}$} (\Eref{eq:zerolag3}) from the maps 
after applying the \textit{bias mask}. We assume $\Delta \chi' \approx$ the width of the photo-$z$ bin. $f$ is 
calculated analytically through \Eref{eq:zerolag2}, where we use $\phi'(z) \propto p'_{l}(z)$, the estimated 
normalized redshift distribution from our photo-$z$ code for each lens bin. 

We combine all estimates for 
the same lens bin $i$ through \Eref{eq:weight_mean} and \Eref{eq:weight_std}, where the covariance between 
the different measurements is estimated using 20 JK samples defined with a ``k-mean'' 
algorithm \citep{MacQueen1967}. The k-mean method splits a set of numbers (center coordinate of 
pixels in our case) into several groups of numbers. The split is made so that the numbers in each group is closest 
to the mean of them. In our analysis it effectively divides our map into areas of nearly equal area, which we use 
as our JK regions. The different JK samples are slightly correlated due to the smoothing 
process. We estimate the effect of this smoothing on the error bars by comparing the JK error bars on the zero-lag 
auto correlation of a random map (with the same size of the data) before and after applying the smoothing. 
For 20 JK samples, this is a $\sim$10\% effect on the error bars, which we will incorporate in the data measurements. 
\chihway{We have also verified that the results are robust to the number of JK samples used.
The above procedure is applied to the data and the simulations using the same analysis pipeline.} 

\chihway{As hinted in \Sref{sec:multiple_bin}, the error bars from JK-resampling do not fully account for the 
uncertainties from cosmic variance. A more complete account for the uncertainty is to measure $\bar{\mu}_{i}$ 
for a large number of simulations that are closely matched to the data. We compare in \Sref{sec:sim_test} the 
resulting error estimation with and without including cosmic variance. 

\subsection{Linear fit}
\label{sec:fit}

In the final step of our analysis, we fit a simple linear model of galaxy bias to the data. To do this, we take into 
account the full covariance between the $\bar{\mu}_{i}$ measurements in different redshift bins, which we 
estimate through simulations. In particular, we use a least-square approach similar to \Eref{eq:weight_mean} 
and consider a linear model for the inverse galaxy bias in the following form
\begin{equation}
D = \boldsymbol{\bar{\mu}} Z,
\end{equation}
where $D = \{\bar{\mu_{i}}\}$ is now the vector containing the measured inverse galaxy bias in each lens redshift 
bin, $\boldsymbol{\bar{\mu}} = \{\bar{\mu}^{0} \; \;\bar{\mu}^{1}\}$ is the vector composed of the two coefficients 
for the linear fit, and $Z = \left( \begin{array}{c}
1 \\ \bar{z}_{i} \end{array} \right) $ is a 2D matrix with the first row being 1 and the second row containing the 
mean redshift of each lens bin. The least-square fit 
to this model and the errors on the fit then becomes
\begin{equation}
\boldsymbol{\bar{\mu}} = Z^{T}C^{-1}D[Z^{T}C^{-1}Z]^{-1},
\end{equation} 
\begin{equation}
\sigma(\boldsymbol{\bar{\mu}})^{2} = (Z^{T}C^{-1}Z)^{-1},
\end{equation} 
where 
\begin{equation}
C^{-1} = \tau \: \mathrm{Cov}^{-1}[D].  %covariance matrix
\end{equation}
Here 
$\tau=(N-\nu-2)/(N-1)$ corrects for the bias in the inverse covariance matrix due to the finite number of 
simulations \citep{Hartlap2007}, where $N$ is the number of simulation samples, and $\nu$ is the 
dimension of $C$. In \Sref{sec:data_measurement}, we only use the four lower redshift bin for the linear 
fit, as the highest redshift bin is unstable and noisy, so $\nu=4$ in our final measurement for the data.}

\begin{figure*}
  \begin{center}
  \includegraphics[scale=0.55]{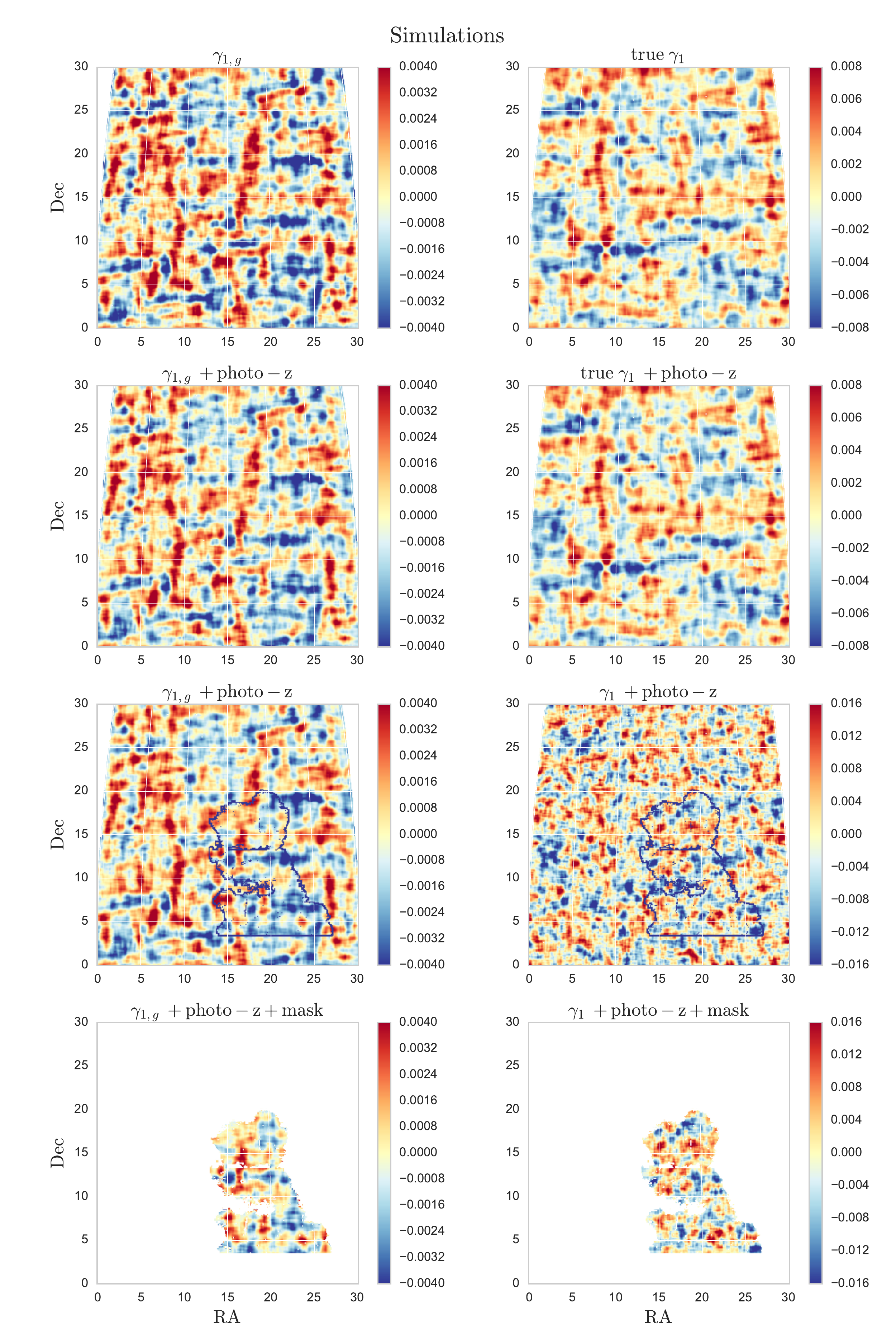} 
  \caption{Example of simulation maps used in this work. The left column show $\gamma_{1,g}$ maps and the right 
  column show $\gamma_{1}$ maps. This $\gamma_{1,g}$ maps are generated from the source redshift 
  bin $z$ (or $ z_{\mathrm{mean}}$)$=1.0-1.2$ and the lens redshift bin $z$ (or $ z_{\mathrm{mean}}$)$=0.4-0.6$. The $\gamma_{1}$ maps are generated from the 
  source redshift bin $z$ (or $ z_{\mathrm{mean}}$)$=1.0-1.2$. 
  The galaxy bias for the lens galaxies can be measured by cross-correlating the left and the right column.
  From top to bottom illustrates the different stages of the degradation of the simulations to match the 
  data. The first row shows the $\gamma_{1,g}$ map against the true $\gamma_{1}$ map for the full 30$\times$30 deg$^{2}$ 
  area. The second row shows the same maps with photo-$z$ errors included, slightly smearing out the structures 
  in both maps. 
  The third row shows the same $\gamma_{1,g}$ map as before against the $\gamma_{1}$ that contains shape noise, making 
  the amplitude higher. Finally, the bottom row shows both maps with the SV mask applied, which is also marked in the third row for 
  reference. Note that the color scales on the $\gamma_{1}$ maps is 2 (4) times higher in the upper (lower) two panels than that of 
  the $\gamma_{1,g}$ maps. }
  \label{fig:sim_maps}
  \end{center}
\end{figure*}

\begin{figure}
  \begin{center}
  \includegraphics[scale=0.45]{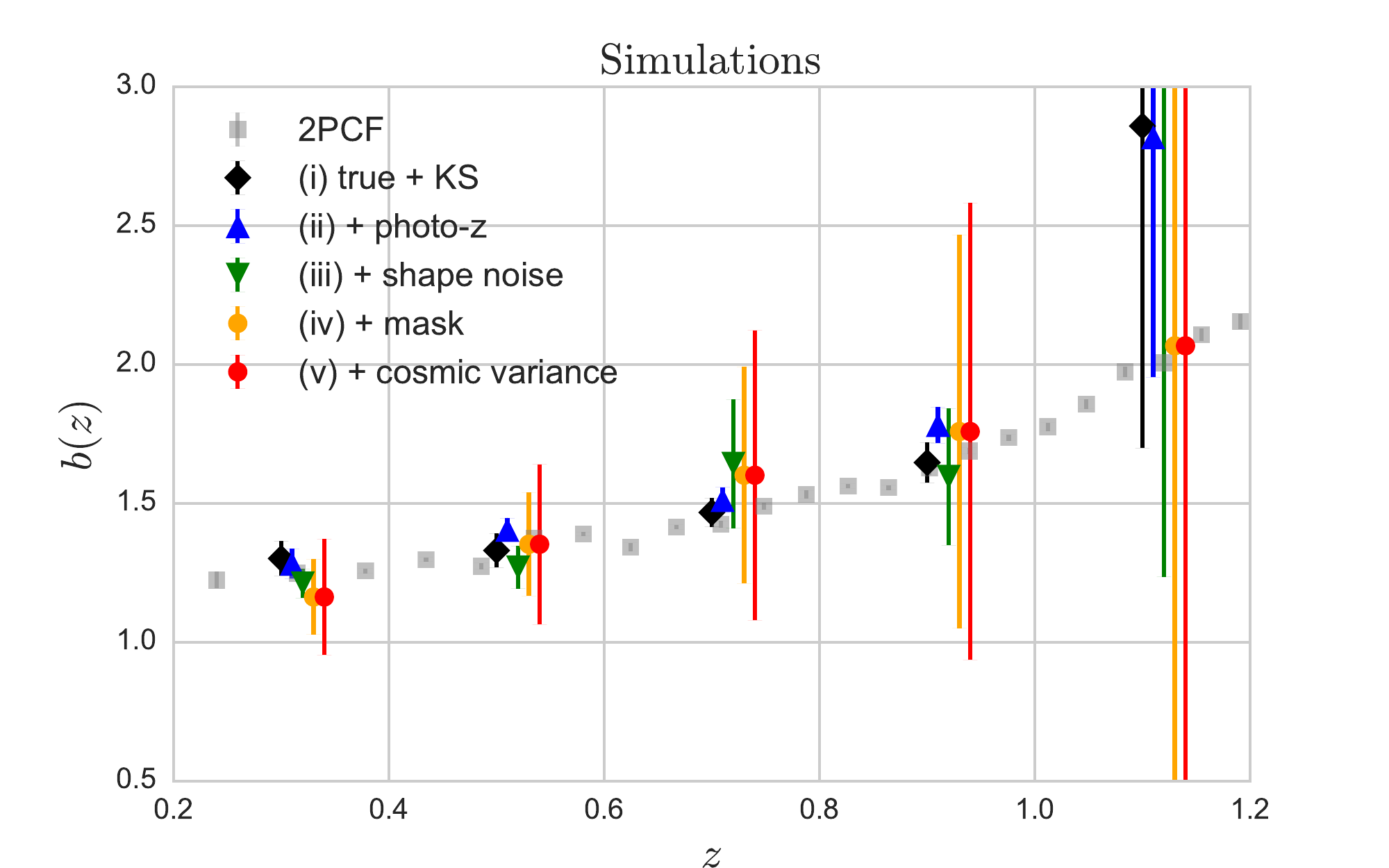} 
  \caption{\chihway{Redshift-dependent galaxy bias measured from simulations with different levels of degradation from the ideal 
  scenario tested in Paper I. The grey  line shows the bias from the 2PCF measurement, which we take as ``truth''. The black, blue, 
  green, orange and red points corresponding to the steps (i), (ii), (iii), (iv) and (v) in \Sref{sec:sim_test}, respectively. The error 
  bars in (i) (ii) and (iii) correspond to the JK error bars (\Eref{eq:weight_std}). The error bars for (iv) is the mean JK error bars for 
  1,200 simulations while the error bars for (v) is the standard deviation of the measurements of 1,200 simulations.}}
  \label{fig:sim_degrade}
  \end{center}
\end{figure}

\subsection{Simulation tests}
\label{sec:sim_test}

Following the procedure outlined above, we present here the result of the redshift-dependent 
galaxy bias measurements from the MICE simulation. We start from an ideal setup in the simulations that 
is very close to that used in Paper I and gradually degrade the simulations until they match our data. Below we 
list the series of steps we take:

\begin{enumerate}
\item use the full area ($\sim900$ deg$^{2}$) with the true $\gvec$ maps
\item repeat above with photo-$z$ errors included
\item repeat above with shape noise included 
\item repeat above with SV mask applied 
\item repeat above with 12 different SV-like areas on the sky, and vary the shape noise 100 times for each
\end{enumerate}
\Fref{fig:sim_maps} illustrates an example of how the $\gamma_{1,g}$ and $\gamma_{1}$ maps degrade over these tests. 
The left column shows the $\gamma_{1,g}$ maps while the right column shows the $\gamma_{1}$ maps. Note that the color 
bars on the upper (lower) two maps in the right panel are 2 (4) times higher compared to the left column. This is to accommodate 
for the large change in scales on the right arising from shape noise in the $\gamma_{1}$ maps. 
The first row corresponds to (i) above, and we can visually see the correspondence of some structures between the two maps. 
Note that the $\gamma_{1,g}$ map only contributes to part of the $\gamma_{1}$ map, which is the reason that we do not expect 
even the true $\gamma_{1,g}$ and $\gamma_{1}$ maps to agree perfectly. 
The second row shows the map with photo-$z$ errors included, corresponding to the step (ii). We find that the real structures in the maps are 
smoothed by the photo-$z$ uncertainties, lowering the amplitude of the map. The smoothing from the photo-$z$ is more visible in 
the $\gamma_{1,g}$ map, since the $\gamma_{1}$ map probes an integrated effect and is less affected by photo-$z$ errors.
The third row shows what happens when shape noise is included, 
which corresponds to the step (iii) above. We find the structures in the $\gamma_{1}$ map becomes barely visible in the presence 
of noise, with the amplitude much higher than the noiseless case as expected. 
The bottom row 
corresponds to the step (iv) above, where the SV mask is applied to both maps. For the $\gamma_{1}$ map this is merely a 
decrease in the area. But for the $\gamma_{1,g}$ map, this also affects the conversion from $\kappa_{g}$ to $\gvec_{g}$, 
causing edge effects in the $\gamma_{1,g}$ map which are visible in the bottom-left map in \Fref{fig:sim_maps}. Step (v) is achieved 
by moving the mask around and drawing different random realizations of shape noise for the source galaxies.

With all maps generated, we then calculate the redshift-dependent galaxy bias following \Eref{eq:weight_mean} 
and \Eref{eq:weight_std} for each 
of the steps from (i) to (v). In \Fref{fig:sim_degrade} we show the result for the different stages, overlaid with the 
bias from the 2PCF measurement described in Paper I. In step (i), our measurements recover the 2PCF 
estimates, confirming the results in Paper I, that we can indeed measure the redshift-dependent bias using this method 
under appropriate settings. Our error bars are smaller than that in Paper I, which is due to the fact that we have combined 
measurements from several source bins, \chihway{and that we estimate inverse-bias instead of bias directly}. Since the only 
difference between this test and the test in Paper I is the inclusion of 
the KS conversion, we have also shown that the KS conversion in the noiseless case does not introduce significant problems 
in our measurements. 
The error bars on the highest redshift bin is large due to the small number of source and lens galaxies. In step (ii), we 
introduce photo-$z$ errors. We find that the photo-$z$ errors do not affect our measurements within the measurement 
uncertainties. \chihway{In step (iii), the error bars increase due to the presence of shape noise. In (iv), we apply the SV 
mask, making the result much noisier due to the smaller area. We repeat this step on 12 different SV-like areas in a 
larger (30$\times$90 deg$^{2}$) simulation area and vary the shape noise realisation 100 times for each area. The 
orange points in \Fref{fig:sim_degrade} shows the average measurement and JK error bars of these 1,200 simulations. 
We find that albeit the large uncertainties, our method indeed gives an unbiased estimation 
of the redshift-dependent of bias which is consistent with the 2PCF estimations. 
In (v), we account for the additional uncertainty in our measurements due to cosmic variance. The red points are 
the same as the orange points, except that the error bars are estimated from the standard deviation of the 1,200 measurements 
in the simulations. We find that the contribution to the uncertainties from cosmic variance can be important especially at 
low redshift.}

\chihway{With the series of simulation tests above, we have shown that our measurement method itself is well 
grounded, but the presence of measurement effects and noise can introduces large uncertainties in the results.
In the next section, we continue with the same measurement on DES SV data and will use the full simulation 
covariance derived in this section for the final fitting process. We believe the simulation covariance matrix captures 
the dominant sources of uncertainties in the problem.} 

\begin{table*}
\begin{center}
\caption{Bias measurement and 1$\sigma$ error bars from DES SV using the method tested in this work, with all possible lens-source 
combinations. We also compare here our main measurements with that using alternative shear and photo-$z$ 
catalogs. Finally we compare our results with other measurement methods carried out on the same data set. 
The C16 estimates are from Tables 3 in that paper, while the G16 estimates are from Table 2 in that paper. }
\begin{tabular}{ccccc}
                            & \multicolumn{4}{c}{Lens redshift ($ z_{\mathrm{mean}}$)} \\ 
                           &  $0.2-0.4$ & $0.4-0.6$ & $0.6-0.8$ & $0.8-1.0$  \\ \hline 
\textbf{This work (\textsc{ngmix}+\textsc{Skynet})  }  &  $\boldsymbol{\btwo}$ & $\boldsymbol{\bthree}$ 
& $\boldsymbol{\bfour}$ & $\boldsymbol{\bfive}$ \\ 
This work (\textsc{im3shape}+\textsc{Skynet})  &1.21$\pm$0.25 & 1.12$\pm$0.24 & 0.90$\pm$0.19 & 0.91$\pm$0.28  \\   
This work (\textsc{ngmix}+\textsc{TPZ})     &1.23$\pm$0.23 & 1.07$\pm$0.18 & 1.39$\pm$0.40 & 1.29$\pm$0.44   \\  
This work (\textsc{ngmix}+\textsc{BPZ})    &0.84$\pm$0.11  & 1.00 $\pm$0.16 & 1.13$\pm$0.26 & 0.95$\pm$0.24  \\ \hline  
\citet{Crocce2016}    & 1.07 $\pm$0.08 & 1.24$\pm$0.04 & 1.34$\pm$0.05 & 1.56$\pm$0.03  \\  
\citet{Giannantonio2016}   & 0.57 $\pm$0.25 & 0.91$\pm$0.22 & 0.68$\pm$0.28 & 1.02$\pm$0.31   \\  
\end{tabular}
\label{tab:bias_sv}
\end{center}
\end{table*}

\begin{figure*}
  \begin{center}
  \includegraphics[scale=0.48]{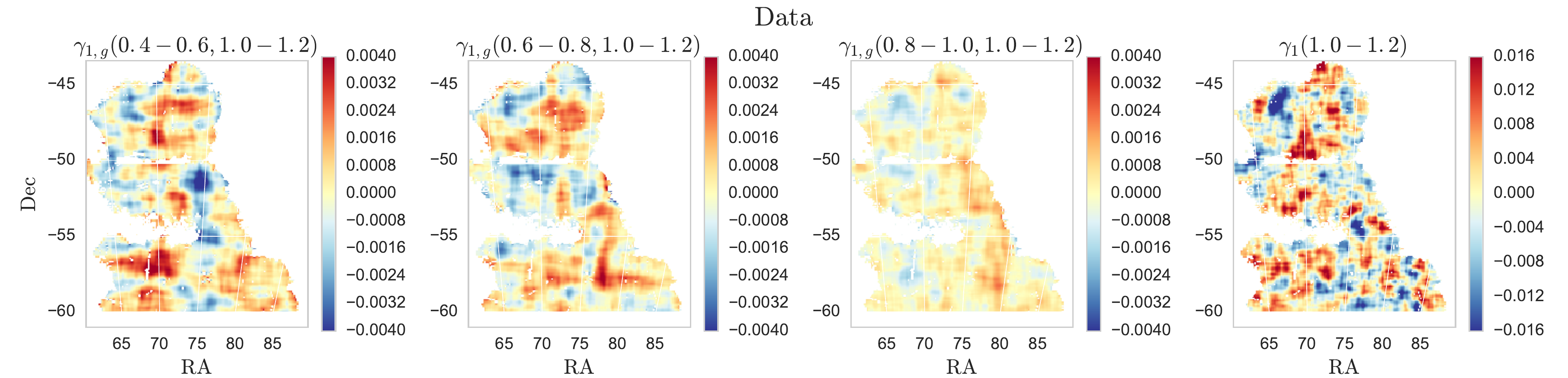} 
  \caption{Example of maps from DES SV data. The right-most panel shows the $\gamma_{1}$ map generated from the 
  source redshift bin $ z_{\mathrm{mean}}=1.0-1.2$, while the other panels show the $\gamma_{1,g}$ maps generated for the 
  source redshift bin $ z_{\mathrm{mean}}=1.0-1.2$ and for different lens redshifts (left: $ z_{\mathrm{mean}}=0.4-0.6$, middle: 
  $ z_{\mathrm{mean}}=0.6-0.8$, right: $ z_{\mathrm{mean}}=0.8-1.0$). The title in each panel for $\gamma_{1,g}$ indicate the lens 
  and source redshift, while the title for $\gamma_{1}$ indicates the source redshift. 
  Note that the color bars are in different ranges, but are matched to the simulation plot in \Fref{fig:sim_maps}. 
  In addition, the left-most and the right-most panels correspond to the bottom row of that figure.}
  \label{fig:data_map}
  \end{center}
\end{figure*}

\subsection{Redshift-dependent galaxy bias of DES SV data}
\label{sec:data_measurement}

We now continue to measure redshift-dependent galaxy bias with the DES SV data using the same procedure 
as in the simulations. \Fref{fig:data_map} shows some examples of the maps. The right-most panel shows the 
$\gamma_{1}$ map at redshift bin $ z_{\mathrm{mean}}=1.0-1.2$, while the rest of the maps are the $\gamma_{1,g}$ maps 
at different redshift bin evaluated for this $\gamma_{1}$ map. We see the effect of the lensing kernel clearly: the 
left-most panel is at the peak of the lensing kernel, giving it a higher weight compared to the other lens bins. We 
also see correlations between $\gamma_{1,g}$ maps at different redshift bins. This is a result of the photo-$z$ 
contamination. 

In \Fref{fig:bias_data} we show the galaxy bias measurement for our magnitude-limited galaxy sample from DES 
SV together with two other independent measurements with the same galaxy sample (discussed in \Sref{sec:comparison}). 
We have excluded the highest redshift bin since with only a small number of source galaxies, the constraining power 
from lensing in that bin is very weak. The black data points show the measurement and uncertainty estimated from 
this work, with a best-fit linear model of: \chihway{$\mu(z) = 1.07^{\pm 0.24}-0.35^{\pm 0.42} z$. The error bars between 
the redshift bins are correlated, and has been taken into account during the fit. 
\Tref{tab:bias_sv} summarizes the results.}

As discussed earlier, our method becomes much less constraining going to higher redshift, as the source galaxies 
become sparse. This is manifested in the increasingly large error bars going to high redshifts. Here we only performed 
a simple linear fit to the data given the large uncertainties in our measurements. In the future, one could extend to  
explore more physically motivated galaxy bias models \citep{Matarrese1997, Clerkin2015}.

Compared with A12, our data set is approximately $\sim$105 times larger, but with a (source) galaxy number 
density $\sim$11.6 times lower. This yields roughly $\sim3$ times lower statistical uncertainty 
in our measurement. Our sample occupies a volume slightly larger than the $0<z<1$ sample in A12. 
Note, however, that due to photo-$z$ uncertainties and the high shape noise per unit area, we expect a slightly 
higher level of systematic uncertainty in our measurement. Since in A12, the emphasis was not on measuring 
linear bias, one should take caution in comparing directly our measurement with A12. But we note that the large 
uncertainties at $z>0.6$ and the weak constraints on the redshift evolution in the galaxy bias is also seen in A12. 
To give competitive constraints on the redshift evolution, higher redshift source planes would be needed.

\subsection{Other systematics test}
\label{sec:systematics}

In \Sref{sec:sim_test}, we have checked for various forms of systematic effects coming from the KS 
conversion, finite area, complicated mask geometry, and photo-$z$ errors. Here we perform three additional 
tests. First, we check that the cross-correlation between the B-mode shear $\gvec_{B}$ and $\gvec_{g}$ is 
small. Next, we check that using the second DES shear pipeline, \textsc{im3shape} gives consistent 
answers with that from \textsc{ngmix}. Finally, we check that using two other photo-$z$ codes also give 
consistent results. These three tests show that there are no significant systematic errors in our 
measurements. 

\subsubsection{B-mode test}

Lensing B-mode refers to the divergent-free piece of the lensing field, which is zero in an ideal, noiseless 
scenario. As a result, B-mode is one of the measures for systematic effects in the data. In \citet{Jarvis2015}, 
a large suite of tests have been carried out to ensure that the shear measurements have lower level of 
systematic uncertainties compared to the statistical uncertainties. Nevertheless, here we test in specific the 
B-mode statistics relevant to our measurements. 

We construct a $\gvec_{B}$ field by rotating the shear measurements in our data by 45 degrees, giving:
\begin{equation}
\gvec_{B} = \gamma_{B,1} + i \gamma_{B,2} = -\gamma_{2} + i \gamma_{1}.
\label{eq:gammaB}
\end{equation}
Substituting $\gvec_{B}$ into $\gvec$ in our galaxy bias calculation (\Eref{eq:zerolag3}) gives an analogous 
measurement to $b$, which we will refer to as $b_{B}$. 
Since we expect $\gvec_{B}$ not to correlate well with $\gvec_{g}$, $1/b_{B}$ would ideally go to zero. 
\chihway{In \Fref{fig:bias_data_B}, we show all the $b_{B}$ measurements using both shear component and all lens-source 
combinations. We see that all the data points are consistent with zero at the 
1--2 $\sigma$ level, assuring that the B-modes in the shear measurements are mostly consistent with noise. We 
also show the weighted mean of all the data points and the corresponding B-mode measurements from one of the 
simulation used in \Sref{sec:sim_test} (iv). We see that the level and scatter in the data is compatible with that in 
the simulations.}

\subsubsection{\textsc{im3shape} test}
\label{sec:sys_im3shape}
\begin{figure}
  \begin{center}
  \includegraphics[scale=0.43]{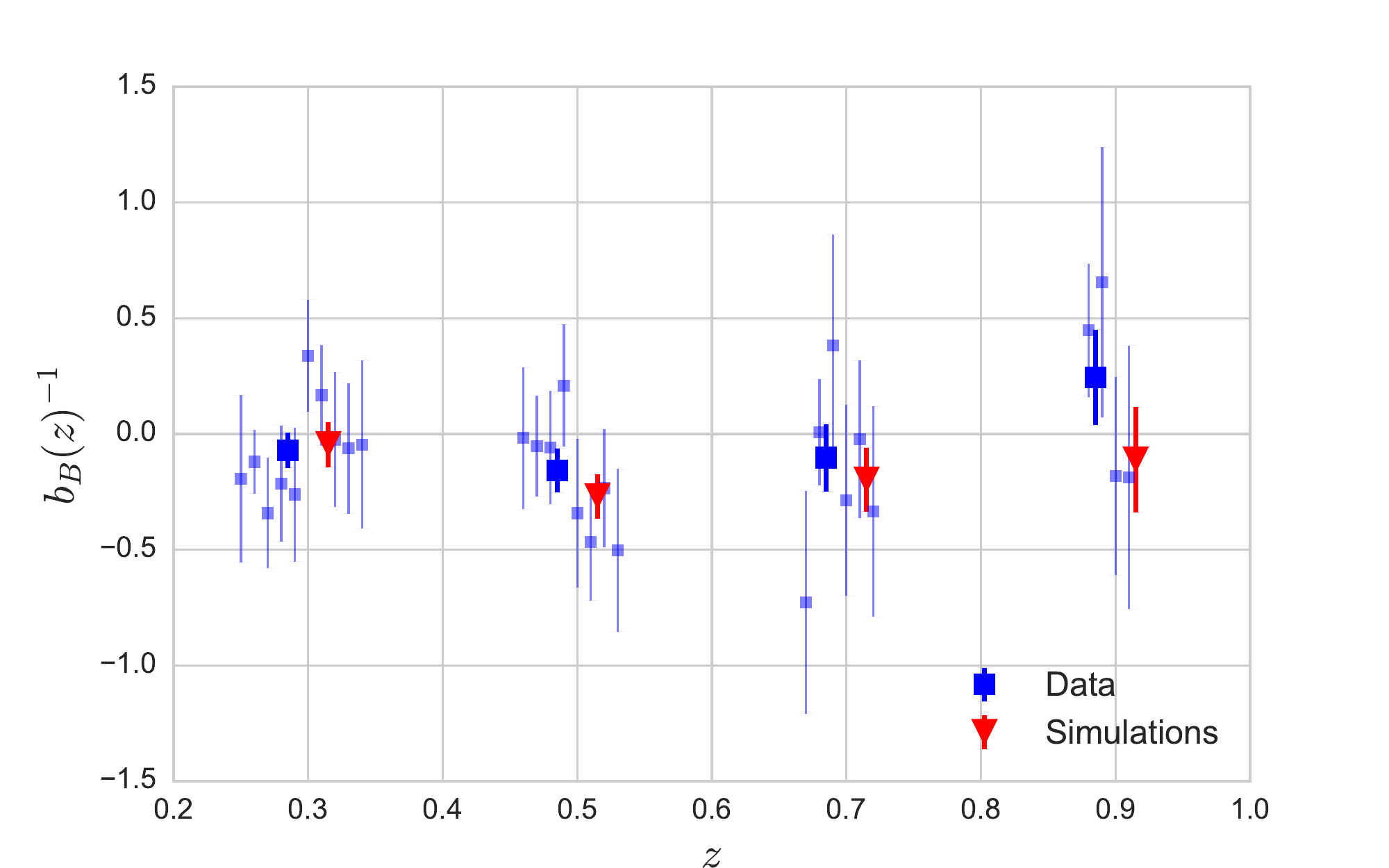} 
  \caption{All $1/b_{B}(z)$ measurements from the B-mode shear and the same $\gvec_{g}$ in our main 
  analysis. Each small blue data point represents a measurement from a combination of lens redshift, source redshift, 
  and shear component. Note that the low redshift bins contain more data points, as there are more source 
  galaxies that can be used for the measurement. The large blue points are the weighted mean of all measurements at 
  the same redshift bin from the DES SV data, while the red points are that from simulations that are well matched to data.  }
  \label{fig:bias_data_B}
  \end{center}
\end{figure}

As described in \Sref{sec:data_shear}, two independent shear catalogs from DES SV were constructed. Here, 
we perform the same measurement in our main analysis using the \textsc{im3shape} catalog. The 
\textsc{im3shape} catalog contains less galaxies, thus the measurements are slightly noisier. The resulting 
redshift-dependent galaxy measurements are shown in \Tref{tab:bias_sv} and are overall slightly higher than 
the \textsc{ngmix} measurements, and there is almost no constraining power on the evolution. 
The best-fit linear bias model is: 
\chihway{$\mu(z) = 0.64^{\pm 0.28}+0.56^{\pm 0.52} z$, which is consistent with the \textsc{ngmix} measurements 
at the 1$\sigma$ level. The B-modes (not shown here) are similar to \Fref{fig:bias_data_B}.}

\subsubsection{Photo-$z$ test}
\label{sec:photoz_test}
As mentioned in \Sref{sec:photoz}, several photo-$z$ catalogs were generated for the DES SV data set and 
shown in \cite{Bonnett2015} to meet the required precision 
and accuracy for the SV data. All above analyzes were carried out with the \textsc{Skynet} photo-$z$ catalog. 
Here we perform the exact same analysis using the other two catalogs: \textsc{BPZ} and 
\textsc{TPZ}. In specific, to be consistent with the other DES SV analyzes \citep{Becker2015, DES2015}, 
we keep the tomographic bins unchanged (binned by \textsc{Skynet} mean redshift), but use the $p(z)$ 
from the different photo-$z$ codes to calculate $f$. The lensing or galaxy maps themselves remain unchanged. 

\Tref{tab:bias_sv} lists the results from the different photo-$z$ catalogs. 
Since \textsc{Skynet} and \textsc{TPZ} are both machine learning codes and 
respond to systematic effects in a similar fashion, while \textsc{BPZ} is a template fitting code, we can 
thus view the difference between the results from \textsc{BPZ} and the others as a rough measure of the potential 
systematic uncertainty in our photo-$z$ algorithm \citep[see also discussion in][]{Bonnett2015}, which is shown 
here to be \chihway{within the $1\sigma$ error bars.}     

\section{Comparison with other measurements}
\label{sec:comparison}

The redshift-dependent galaxy bias has been measured on the same data set using other approaches. 
Here we compare our result with two other measurements -- galaxy clustering 
\citep[][hereafter C16]{Crocce2016} and cross-correlation of galaxies and CMB lensing 
\citep[][hereafter G16]{Giannantonio2016}.We note that both these analyzes assumed the 
most recent \textit{Planck} cosmological parameters \citep{PlanckCollaboration2014}, which is slightly different 
from our assumptions (see \Sref{sec:sims}). But since our measurement depends very weakly on the 
assumption of cosmological parameters (as discussed in \Sref{sec:bias_cl_wl}), the stronger cosmology 
dependencies come from the cosmological parameters assumed in C16 and G16, which are 
known well within our measurement uncertainties. We also note that the results we quote in 
\Tref{tab:bias_sv} are based on the photo-$z$ code \textsc{TPZ}, which means our redshift binning 
is not completely identical to theirs. 

\begin{figure}
  \begin{center}
  \includegraphics[scale=0.43]{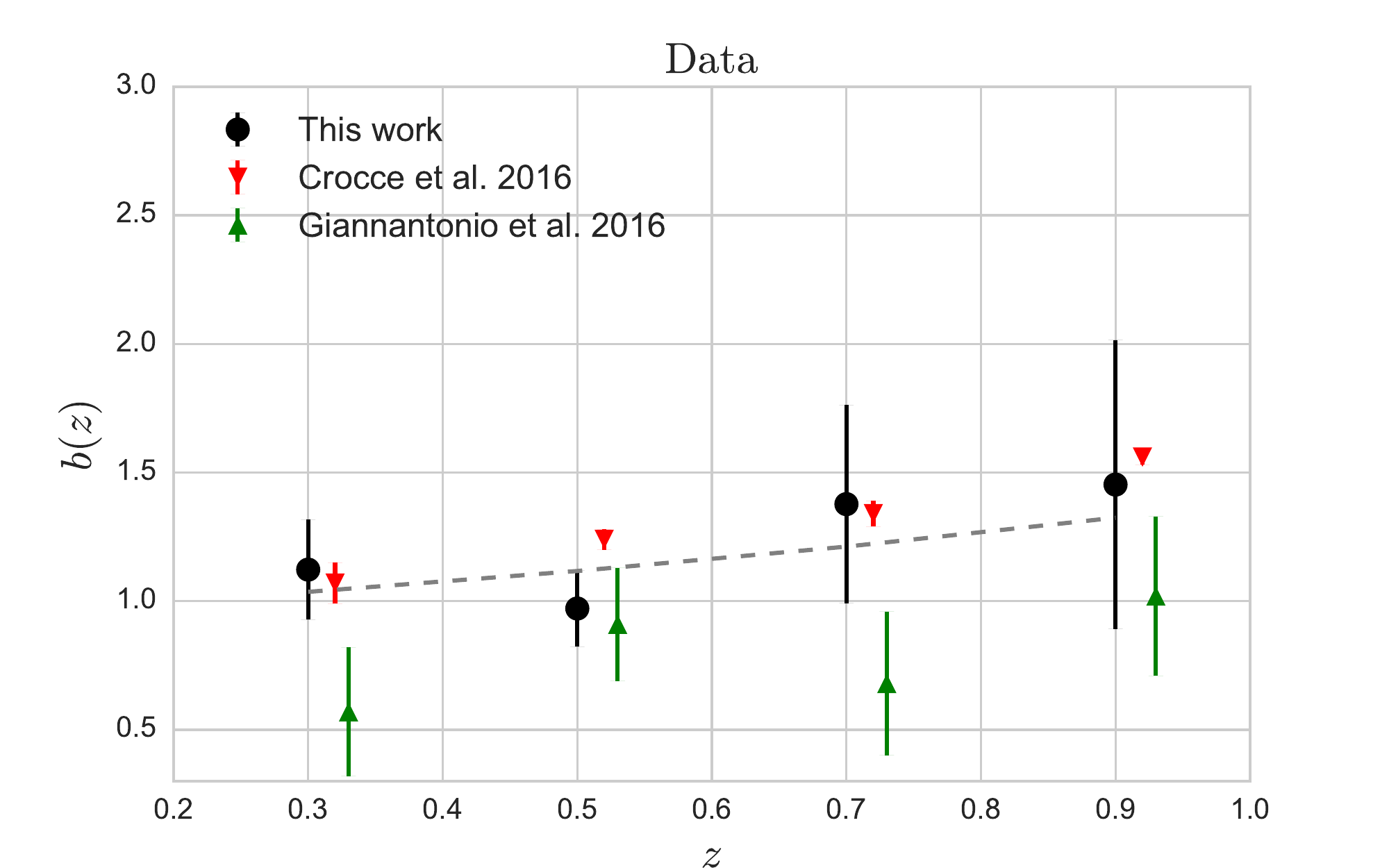} 
  \caption{Redshift-dependent bias measured from the DES SV data. The black data points show the 
  result from this work. The red and green 
  points show the measurements on the same galaxy sample with different methods. 
    The grey dashed line is the best fit to the black data points.}
  \label{fig:bias_data}
  \end{center}
\end{figure}

\subsection{Bias measurement from galaxy clustering}

In C16, galaxy bias was estimated through the ratio between the projected galaxy angular correlation 
function (2PCF) in a given redshift bin and an analytical dark matter angular correlation function predicted at the 
same redshift. The latter includes both linear and nonlinear dark matter clustering derived from \textsc{CAMB} 
\citep{Lewis2000} assuming a set of cosmological parameters. In C16, a flat $\Lambda$CDM+$\nu$ 
cosmological model based on \textit{Planck} 2013 + \textit{WMAP} polarization + ACT/SPT + BAO was used.
The results in C16 as listed in \Tref{tab:bias_sv} were shown to be consistent with the independent 
measurement from the CFHTLS \citep{Coupon2012}.

Compared to C16, our work aims to measure directly the local galaxy bias (\Eref{eq:bias}) instead of the 
galaxy bias defined through the 2PCF (\Eref{eq:xi}). Although the two measurements agree in the linear 
regime where this work is based 
on, comparing the measurements on smaller scales will provide further insight to these galaxy bias models.
Our method is less sensitive to assumptions of cosmological parameters compared to the 2PCF method. In 
particular, it does not depend strongly on $\sigma_{8}$, which breaks the degeneracy between 
$\sigma_{8}$ and the measured galaxy bias $b$ in other measurement methods. Finally, since our 
measurement is a cross-correlation method (compared to C16, an auto-correlation method), it suffers less 
from systematic effects that only contaminate either the lens or the source sample. On the other hand, 
however, lensing measurements are intrinsically noisy and the conversion between shear and convergence 
is not well behaved in the presence of noise and complicated masking. In addition, we only considered 
a one-point estimate (zero-lag correlation), which contains less information compared to the full 2PCF 
functions. All these effects result in much less constraining power in our measurements. 

\chihway{As shown in \Fref{fig:bias_data} and listed in \Tref{tab:bias_sv}, our measurements and C16 agree 
very well except for the redshift bin $z=0.4-0.6$ (slightly more than 1$\sigma$ discrepancy).} We note, 
however, both C16 and our work may not have included the complete allocation of systematic errors (especially 
those coming from the photo-$z$ uncertainties), which could introduce some of the discrepancies. 

\subsection{Bias measurement from cross-correlation of galaxies and CMB lensing}

In G16, galaxy bias is estimated by the ratio between the galaxy-CMB convergence cross-correlation 
and an analytical prediction of the dark matter-CMB convergence cross-correlation, both calculated 
through the 2PCF (and also in harmonic space through the power spectrum). Since the lensing efficiency 
kernel of the CMB is very broad and the CMB lensing 
maps are typically noisy, this method has less constraining power than C16. However, by using an 
independent external data, the CMB lensing maps from the South Pole Telescope and the \textit{Planck} 
satellite, this measurement serves as a good cross check for possible systematic effects in the DES data.

In calculating the theoretical dark matter-CMB convergence cross-correlation, G16 also assumed a 
fixed cosmology and derived all predictions using \textsc{CAMB}. The $\sigma_{8}$-$b$ degeneracy is thus 
also present in their analysis. We note, however, that one could apply our method to the CMB lensing 
data and avoid this dependency. In our framework, the CMB lensing plane will serve as an additional 
source plane at redshift $\sim$1100. We defer this option to future work. 

The results from G16 are shown in \Fref{fig:bias_data} and listed in \Tref{tab:bias_sv}. These 
results come from the ratio between the measured and the predicted power spectrum, which suffers 
less from non-linear effects compared to the measurement in real space (2PCF).
\chihway{We find that G16 is systematically lower than our measurement at the 1-2$\sigma$ level for all
redshift bins.} G16 also has more constraining power at high redshift compared to our results, as expected.
Possible reasons for the discrepancy at low redshift include systematic errors (in e.g. the photo-$z$ estimation) 
that are not included in either C16, G16 or this work. In addition, the redshift bins are significantly covariant, making 
the overall discrepancy less significant. Finally, the scales used in the three studies are slightly different.
We refer the readers to G16 for more discussion of this discrepancy.

\section{Conclusion}
\label{sec:conclusion}

In this paper, we present a measurement of redshift-dependent bias using a novel technique of cross-correlating 
the weak lensing shear maps and the galaxy density maps. The method serves as an alternative measurement to 
the more conventional techniques such as 2-point galaxy clustering, and is relatively insensitive to the assumed 
cosmological parameters. The method was first developed in \citet{Amara2012} 
and later tested more rigorously with simulations in a companion paper \citep[][Paper I]{Pujol2016}. Here we 
extend the method and apply it on  
wide-field photometric galaxy survey data for the first time. We measure the galaxy bias for a magnitude-limited 
galaxy sample in the Dark Energy Survey (DES) Science Verification (SV) data. 

Following from Paper I, we carry out a series of simulation tests which incorporate step-by-step realistic effects in 
our data including shape noise, photo-$z$ errors and masking. In each step, we investigate the errors introduced 
in our estimation of galaxy bias. \chihway{We find that shape noise and cosmic variance are the main sources of uncertainties, 
while the photo-$z$ affects the measurements in a predictable way if the characteristics of the photo-$z$ 
uncertainties are well understood. As the measurement itself 
is very noisy, simulation tests where we know the ``truth'' provide a good anchor for building the analysis pipeline.  }

In our main analysis, we measure the galaxy bias with a $18<i<22.5$ magnitude-limited galaxy sample in 4 
tomographic redshift bins to be 
\chihway{
$\btwo$ ($z=0.2-0.4$), $\bthree$ ($z=0.4-0.6$), 
$\bfour$ ($z=0.6-0.8$), and $\bfive$ ($z=0.8-1.0$).}  
Measurements from higher redshifts are too noisy to be constraining. The best-fit linear model gives: 
\chihway{$b^{-1}(z) =\mu(z)= 1.07^{\pm 0.24}-0.35^{\pm 0.42} z$}. The results are consistent between different 
shear and photo-$z$ catalogs.

The galaxy bias of this same galaxy sample has 
also been measured with two other techniques described in \citet{Crocce2016} and \citet{Giannantonio2016}. The  
three measurements agree at the 1--2$\sigma$ level at all four redshift bins, though the results from 
\citet{Giannantonio2016} are 
systematically lower than our measurements. We note that our method is more constraining at 
low redshift regions where there are more source galaxies behind the lens galaxies. As pointed out in 
\citet{Amara2012}, to constrain the evolution of galaxy bias, our current data set may not be optimal. A more 
efficient configuration would be combining a wide, shallow data set with a narrow, deep field. We plan on 
exploring these possibilities in the future.
\chihway{The main uncertainty in this work comes from the combined effect of masking, shape noise and cosmic 
variance. However, as we 
demonstrated with simulations, moving to the larger sky coverage of the first and second year of DES data 
would reduce this effect significantly. }

We have demonstrated the feasibility and validity of our method for measuring galaxy bias on a wide-field 
photometric data set. Looking forward to the first and second year of DES data ($\sim$2,000 square degrees and 
$\sim1$ magnitude shallower), we expect to explore a variety of other topics using this method with the increased 
statistical power. For example, the same measurement could be carried out on different subsamples of lens galaxies 
(in magnitude, color, galaxy type etc.) and gain insight into the different clustering properties for different galaxy 
populations. Also, one can extend the measurement into the non-linear regime and measure the scale-dependencies 
of the galaxy bias. Finally, it would be interesting to compare the measurement from the 2PCF method and our method 
(which is a measure of local bias) on different scales to further understand the connections between the two galaxy 
bias models. 

\section*{Acknowledgement} 

We thank Marc Manera, Donnacha Kirk, Andrina Nicola, Sebastian Seehars for useful discussion and feedback. 
CC, AA, AR and TK are supported by the Swiss National Science Foundation grants 200021-149442 and 
200021-143906. AP was supported by beca FI and 2009-SGR-1398 from Generalitat de Catalunya 
and project AYA2012-39620 from MICINN. JZ and SB acknowledge support from the European Research 
Council in the form of a Starting Grant with number 240672.

We are grateful for the extraordinary contributions of our CTIO colleagues and the DECam Construction, 
Commissioning and Science Verification teams in achieving the excellent instrument and telescope conditions 
that have made this work possible.  The success of this project also relies critically on the expertise and 
dedication of the DES Data Management group.

Funding for the DES Projects has been provided by the U.S. Department of Energy, the U.S. National 
Science Foundation, the Ministry of Science and Education of Spain, the Science and Technology Facilities 
Council of the United Kingdom, the Higher Education Funding Council for England, the National Center for 
Supercomputing Applications at the University of Illinois at Urbana-Champaign, the Kavli Institute of 
Cosmological Physics at the University of Chicago, the Center for Cosmology and Astro-Particle Physics at 
the Ohio State University, the Mitchell Institute for Fundamental Physics and Astronomy at Texas A\&M University, 
Financiadora de Estudos e Projetos, Funda{\c c}{\~a}o Carlos Chagas Filho de Amparo {\`a} Pesquisa do Estado 
do Rio de Janeiro, Conselho Nacional de Desenvolvimento Cient{\'i}fico e Tecnol{\'o}gico and the Minist{\'e}rio da 
Ci{\^e}ncia, Tecnologia e Inova{\c c}{\~a}o, the Deutsche Forschungsgemeinschaft and the Collaborating 
Institutions in the Dark Energy Survey. 

The Collaborating Institutions are Argonne National Laboratory, the University of California at Santa Cruz, the 
University of Cambridge, Centro de Investigaciones Energ{\'e}ticas, Medioambientales y Tecnol{\'o}gicas-Madrid, 
the University of Chicago, University College London, the DES-Brazil Consortium, the University of Edinburgh, 
the Eidgen{\"o}ssische Technische Hochschule (ETH) Z{\"u}rich, 
Fermi National Accelerator Laboratory, the University of Illinois at Urbana-Champaign, the Institut de Ci{\`e}ncies 
de l'Espai (IEEC/CSIC), the Institut de F{\'i}sica d'Altes Energies, Lawrence Berkeley National Laboratory, the 
Ludwig-Maximilians Universit{\"a}t M{\"u}nchen and the associated Excellence Cluster Universe, the University of 
Michigan, the National Optical Astronomy Observatory, the University of Nottingham, The Ohio State University, 
the University of Pennsylvania, the University of Portsmouth, 
SLAC National Accelerator Laboratory, Stanford University, the University of Sussex, and Texas A\&M University.

The DES data management system is supported by the National Science Foundation under Grant Number AST-1138766.
The DES participants from Spanish institutions are partially supported by MINECO under grants AYA2012-39559, 
ESP2013-48274, FPA2013-47986, and Centro de Excelencia Severo Ochoa SEV-2012-0234.
Research leading to these results has received funding from the European Research Council under the European Union's 
Seventh Framework Programme (FP7/2007-2013) including ERC grant agreements 240672, 291329, and 306478.
%\bibliographystyle{mn2e}
%\bibliography{bias_des.bib}

\begin{figure}
  \begin{center}
  \includegraphics[scale=0.43]{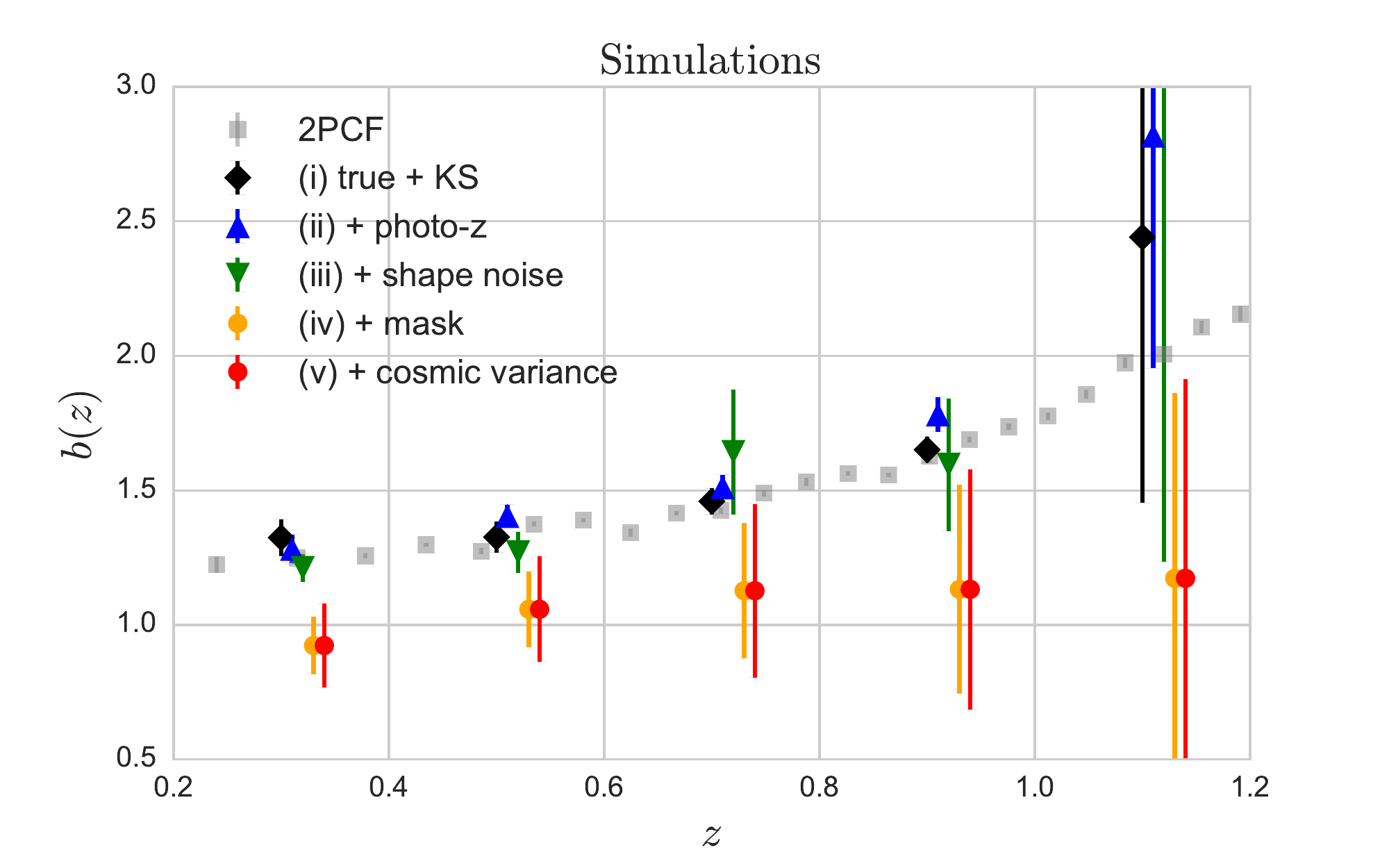} 
  \caption{Same as \Fref{fig:sim_degrade}, but using $b$ as the estimator.}
  \label{fig:sim_degrade_b}
  \end{center}
\end{figure}

\begin{figure}
  \begin{center}
  \includegraphics[scale=0.43]{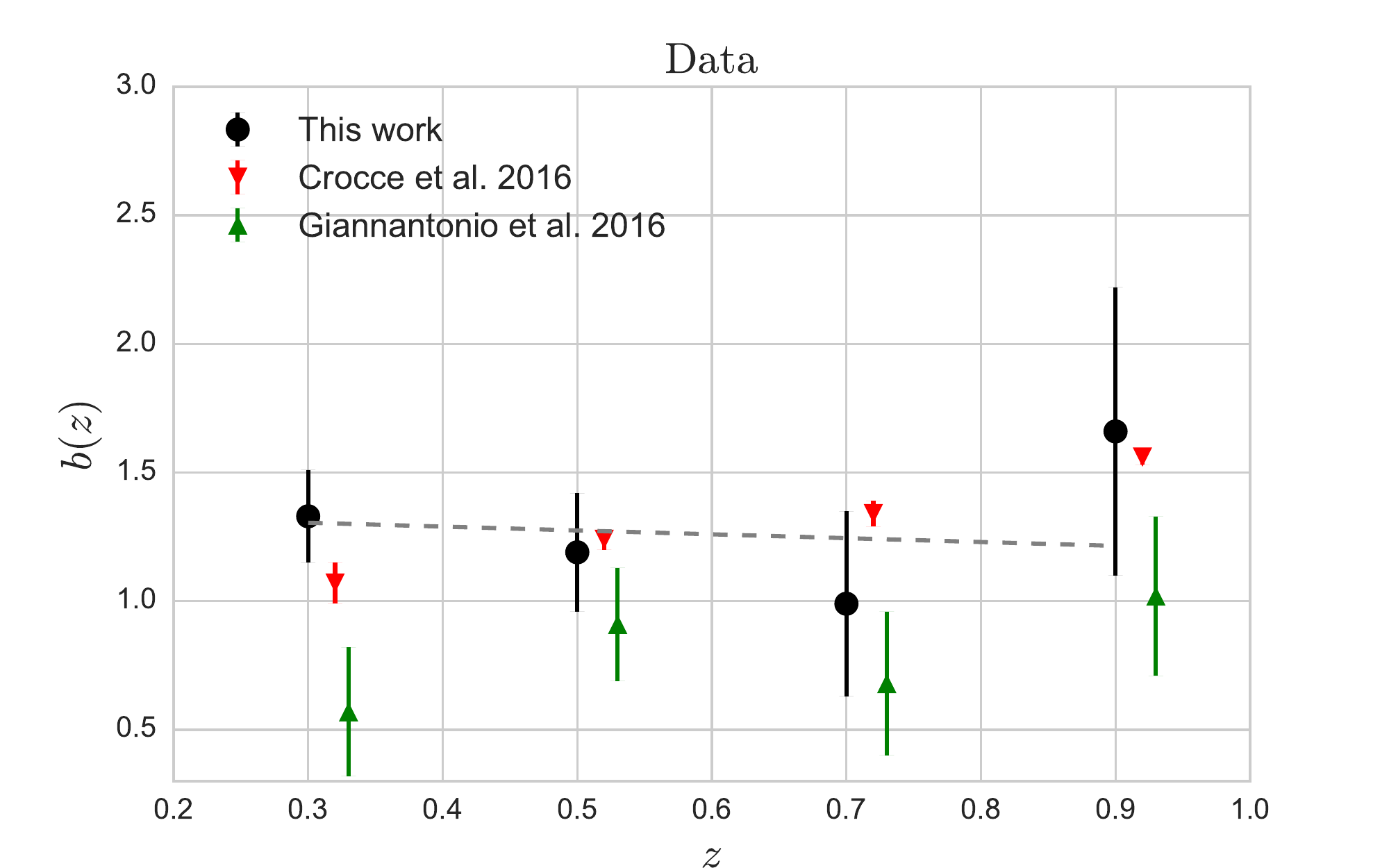} 
  \caption{Same as \Fref{fig:bias_data}, but using $b$ as the estimator.}
  \label{fig:bias_data_b}
  \end{center}
\end{figure}

\appendix
\chihway{
\section{Choice of estimator}
\label{sec:estimator}

In our main analysis, we use the inverse-galaxy bias $\mu=1/b$ as our main estimator instead of 
estimating galaxy bias directly. In this appendix we show the effect of using $b$ as the estimator. The origin 
of the difference comes from the fact that when combining the multiple measurements in the same lens bin, 
we use the least-square formalism \Eref{eq:weight_mean}, which relies on the covariance matrix $C_{i}$ 
estimated through JK resampling. This covariance matrix can become ill-behaved depending on the estimator 
used. In our case, the denominator of $b$ (the inverse of \Eref{eq:zerolag3})
can become close to zero, which makes the inversion of the covariance matrix unstable. We find that 
this introduces a bias in our final result, which will need to be calibrated.  

In \Fref{fig:sim_degrade_b} we show the equivalent of \Fref{fig:sim_degrade} using $b$ as an estimator 
instead of $\mu$. As the distribution of $b$ estimated through the simulations have large outliers, we 
exclude simulations with bias estimates below 0 and above 5. We find that the main difference between 
\Fref{fig:sim_degrade} and \Fref{fig:sim_degrade_b} is in the orange and red points, where all the observational 
effects are included. For the less noisy scenarios (i)(ii) and (iii), the change is very minor. This is because the 
effect is more manifested when the measurements are noisy. The final (red) points in \Fref{fig:sim_degrade_b} 
is biased from the ``truth'' by $\Delta b$ due to the matrix inversion discussed above. 

Once we calibrate $\Delta b$ from these simulations and apply it to the data measurements, we have 
\Fref{fig:bias_data_b}, which is the equivalent of \Fref{fig:bias_data} but using $b$ as an estimator instead of 
$\mu$. We find that after taking into account the bias derived from \Fref{fig:sim_degrade}, the final measurements 
from the data is still consistent with our main analysis in \Fref{fig:bias_data}. Nevertheless, as using $b$ relies 
heavily of the quality of the simulations and the outlier-rejection described above is not entirely objective, we 
choose to use the estimator $\mu$ instead.}

\section*{Affiliations}
{\small
$^{1}$\ethz \\
$^{2}$\ieec \\
$^{3}$\ports \\
$^{4}$\stanford \\
$^{5}$\kipac \\
$^{6}$\ifae \\
$^{7}$\cambridge \\
$^{8}$\cambridgekavli \\
$^{9}$\damtp\\
$^{10}$\upenn \\
$^{11}$\ccap \\
$^{12}$\bnl \\
$^{13}$\manchester \\
$^{14}$\anl \\
$^{15}$\ctio \\
$^{16}$\ucl \\
$^{17}$\rhodes \\
$^{18}$\fermilab \\
$^{19}$\paris \\
$^{20}$\sorbonne \\
$^{21}$\slac \\
$^{22}$\lina \\
$^{23}$\on \\
$^{24}$\uiuc \\
$^{25}$\ncsa \\
$^{26}$\southhampton \\
$^{27}$\lmu \\
$^{28}$\cluster \\
$^{29}$\jpl \\
$^{30}$\michiganastro \\
$^{31}$\michigan \\
$^{32}$\kicp \\
$^{33}$\berkeley \\
$^{34}$\lbnl \\
$^{35}$\maxplanck \\
$^{36}$\ohio \\
$^{37}$\aao \\
$^{38}$\texas \\
$^{39}$\fisica \\
$^{40}$\princeton \\
$^{41}$\barcelona \\
$^{42}$\sussex \\
$^{43}$\ciemat \\
}

\end{document}